\documentclass[aps,prl,reprint,superscriptaddress,showpacs,nofootinbib]{revtex4-1}

\usepackage{amsmath}
\usepackage{graphicx}
\usepackage{epstopdf}
\usepackage{latexsym}
\usepackage{amssymb}
\usepackage{hyperref}
\usepackage{comment}
\hypersetup{
    colorlinks=true,
    linkcolor=blue,
    citecolor=blue,
    urlcolor=blue
}
\usepackage{array}
\usepackage{multirow}
\usepackage{enumitem}
\usepackage{float}
\usepackage{url}
\usepackage[thinc]{esdiff}

\begin{document}


\title{Complementary probes of inflationary cosmology}

\author{Jurgen Mifsud}
\email{jmifsud1@sheffield.ac.uk}
\affiliation{Consortium for Fundamental Physics, School of Mathematics and Statistics, University of Sheffield, Hounsfield Road, Sheffield S3 7RH, UK}
\author{Carsten van de Bruck}
\email{c.vandebruck@sheffield.ac.uk}
\affiliation{Consortium for Fundamental Physics, School of Mathematics and Statistics, University of Sheffield, Hounsfield Road, Sheffield S3 7RH, UK}

\date{\today}

\begin{abstract}
A rigorous constraint analysis on cosmic inflation entails several probes which collectively survey an extensive range of energy scales. We complement the cosmic microwave background data with an updated compilation of the cosmic abundance limits of primordial black holes, with which we infer stringent constraints on the runnings of the scalar spectral index. These constraints are notably improved by further including imminent measurements of the cosmic microwave background spectral distortions, clearly illustrating the effectiveness of joint large-scale and small-scale cosmological surveys.
\end{abstract}


\maketitle



Measurements of the anisotropies in the cosmic microwave background (CMB) radiation placed very tight constraints on the parameters of the $\Lambda$CDM model \cite{Ade:2015xua}, and provided evidence that cosmic inflation \cite{Starobinsky:1979ty,Kazanas:1980tx,Guth:1980zm,Linde:1981mu,Sato:1980yn,Albrecht:1982wi} describes well the dynamics of the early Universe (see e.g. Refs. \cite{Martin:2013nzq,Guth:2013sya,Burgess:2013sla,Akrami:2018odb} and references therein). We will be focusing on possible small-scale departures in a generic profile of the primordial curvature power spectrum from the nearly scale-invariant spectrum that is very well-constrained on the large-scales. Such a deviation could be simply addressed by including higher-order parameters of the primordial curvature power spectrum which are often neglected, since, as we will show, the large-scale probes of the Universe are not very sensitive to these parameters. We will demonstrate that by jointly considering large-scale and small-scale data sets, we can place robust constraints on the physics of the early Universe not accessible by any other means.

In the following, we will discuss the cosmological consequences of tiny departures from the Planck energy spectrum of the CMB radiation, leading to spectral distortions \cite{Weymann1966,Zeldovich:1969ff,Sunyaev:1970er,Daly1991,Barrow_Coles1991} in the CMB radiation spectrum. This offers a crucial window on the primordial power spectrum at small-scales, and we explicitly show its impact on current inflationary constraints by adopting forecasted spectral distortion measurements. Furthermore, we will be making use of early Universe constraints arising from the non-detection of primordial black holes (PBHs) \cite{1967SvA....10..602Z,1971MNRAS.152...75H,1974Natur.248...30H,1974MNRAS.168..399C}. We remark that the spatially-flat Friedmann-Lema\^{i}tre-Robertson-Walker metric is assumed.


\textit{CMB spectral distortions.}-- Apart from the cosmic signatures contained within the CMB anisotropies, the energy spectrum of the CMB provides us with invaluable information about the thermal history of the Universe at very early times. There are several physical mechanisms which could lead to an energy release in this primordial era, such as decaying relic particles \cite{Hu:1993gc,Chen:2003gz}, and annihilating particles \cite{Padmanabhan:2005es,Chluba:2009uv}. We will be considering two scenarios that are present in the standard model of cosmology: the adiabatic cooling of electrons and baryons, together with the dissipation of acoustic waves. The latter mechanism arises from the Silk damping \cite{Silk:1967kq} of primordial small-scale fluctuations leading to an energy release in the early Universe. Consequently, inevitable spectral distortions of the CMB spectrum \cite{Weymann1966,Zeldovich:1969ff,Sunyaev:1970er,Daly1991,Barrow_Coles1991} are produced, which are sensitive to the underlying functional form of the primordial power spectrum. In this analysis we will be interested in those modes $\left(50\,\mathrm{Mpc}^{-1}\lesssim k \lesssim 10^4\,\mathrm{Mpc}^{-1}\right)$ which dissipate their energy during the $\mu$-era $\big(5\times10^{4}\lesssim z \lesssim 2\times10^{6}\big)$ producing a small residual chemical potential. As long as Compton scattering is still able to achieve full kinetic equilibrium with electrons, the CMB spectrum is able to regain a Bose-Einstein distribution with chemical potential $\mu$, and occupation number $n_{\mathrm{BE}}^{}(x\equiv h_{Pl}^{}\nu/k_{B}^{}T_{\gamma})=1/(e^{x+\mu}-1)$, with $T_{\gamma}$ denoting the CMB temperature, $k_{B}^{}$ is Boltzmann's constant, and $h_{Pl}^{}$ is Planck's constant. In general, $\mu$ will be frequency-dependent due to the creation of photons at low frequencies, although experiments like the Primordial Inflation Explorer (\textit{PIXIE}) \cite{Kogut:2011xw} are expected to probe the high frequency spectrum $(30\,\mathrm{GHz}\leq\nu\leq6\,\mathrm{THz})$, in which the chemical potential is constant \cite{Chluba:2011hw,Khatri:2012tv}. At higher redshifts $\left(z\gtrsim2\times10^{6}\right)$, double Compton scattering, bremsstrahlung and Compton scattering are so efficient that the thermalization process ensures that for nearly any arbitrary amount of energy injection, no spectral distortion should remain today \cite{Daly1991}. At lower redshifts $\left(z\lesssim10^{4}-10^{5}\right)$, the Compton redistribution of photons over the entire spectrum is too weak to establish a Bose-Einstein spectrum, resulting in a $y$-distortion. Since $y$-type spectral distortions continue to be created throughout the late-time epoch of the Universe, we will marginalize over it in our analysis.

At redshifts well before the recombination era $(z\gtrsim10^{4})$, one can use the tight coupling approximation to compute the energy injection rate from the photon-baryon fluid acoustic wave dissipation, given by \cite{Chluba:2012gq,Khatri:2012rt,Khatri:2013dha}
\begin{equation}
\label{eqn:dQdz}
\diff{Q}{z}\bigg\vert_{\mathrm{ac}}=\frac{9}{4}\diff{k_{D}^{-2}}{z}\int\frac{\mathrm{d}^{3}k}{(2\pi)^{3}}P_{\zeta}(k)\,k^{2}\,e^{-2k^{2}/k_{D}^{2}}\;,
\end{equation}
where the primordial power spectrum is defined by $P_{\zeta}(k)=4/(0.4R_\nu+1.5)\mathcal{P}_{\zeta}\approx1.45\mathcal{P}_{\zeta}$, with $R_\nu=\rho_\nu/(\rho_\gamma+\rho_\nu)\approx0.41$ denoting the contributions of massless neutrinos $(\rho_\nu)$ to the energy density of relativistic species. The phenomenological parametrization of the curvature power spectrum of scalar perturbations in the comoving gauge $\mathcal{P}_{\zeta}$, is given by \cite{Kosowsky:1995aa}
\noindent
\begin{equation}
\label{eqn:P_zeta}
\mathcal{P}_{\zeta}\hspace{-1.5pt}=\hspace{-1.5pt}\frac{2\pi^2}{k^3}\hspace{-.5pt}A_s\hspace{-1.5pt}\left(\frac{k}{k_{0}^{}}\right)^{n_s^{}-1+\frac{\alpha_s^{}}{2}\ln\left[k/k_{0}^{}\right]+\frac{\beta_s^{}}{6}\ln^{2}_{}\left[k/k_{0}^{}\right]+\frac{\gamma_s^{}}{24}\ln^{3}_{}\left[k/k_{0}^{}\right]}\hspace{-2pt},
\end{equation}
where $A_s$ is the scalar amplitude, $n_s$ denotes the scalar spectral index, and $\alpha_s$, $\beta_s$, and $\gamma_s$ are the runnings of the scalar spectral index. We assume that these parameters are all specified at the pivot scale $k_0^{}$. Although one could consider additional higher-order terms in the above exponent, current data sets are unable to place limits on these parameters, and we therefore restrict our analyses up to $\beta_s$.
\begin{figure}[t!]
\begin{center}
\includegraphics[width=0.99\columnwidth]{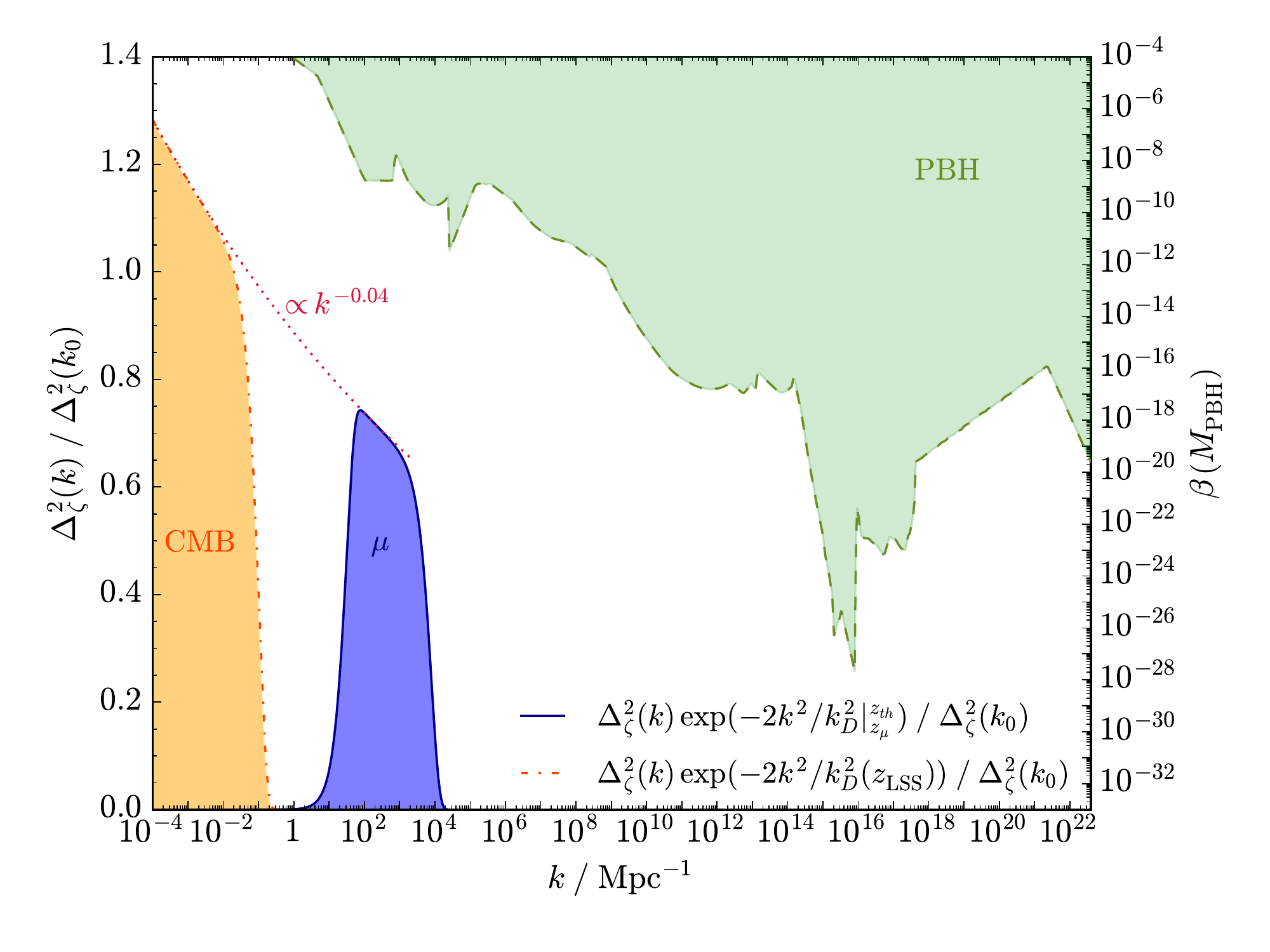}
\caption{We depict the distinct length scales that are probed by the CMB anisotropies, $\mu$-distortion, and PBHs. For the CMB and $\mu$-distortion shaded regions we plot $\Delta_\zeta^2(k)\mathrm{exp}\left(-2k^2/k_D^2\right)/\Delta_\zeta^2(k_0^{})$ at the redshift of the last-scattering surface $z_\mathrm{LSS}=1100$, and its difference between $z_{th}$ and $z_\mu$, respectively. For the PBH shaded region, we illustrate the full compilation of the constraints on $\beta\left(M_\mathrm{PBH}\right)$ (see Eq. (\ref{eqn:beta_PBH})), as reported in Ref. \cite{Mifsud:PBH_PRD}.
\label{fig:mu_window}}
\end{center}
\end{figure}
%
 We further define the dimensionless primordial power spectrum by
\begin{equation}
\label{eqn:dimensionles_P_zeta}
\Delta_\zeta^2(k)=\frac{k^3\mathcal{P}_\zeta(k)}{2\pi^2}\;,
\end{equation}
and the photon damping scale by \cite{Silk:1967kq,Peebles:1970ag,Kaiser:1983,Zaldarriaga:1995gi}
\begin{equation}
k_{D}^{-2}(z)=\int_{z}^{\infty}\mathrm{d}z^{\prime}\frac{c(1+z^{\prime})}{6H(1+R)n_e\sigma_{T}^{}}\left(\frac{R^2}{1+R}+\frac{16}{15}\right)\;,
\end{equation}
where $n_e$ is the number density of electrons, $\sigma_{T}^{}$ is the Thomson scattering cross-section, $c$ is the speed of light, $H$ is the Hubble parameter, and $R=3\rho_{b}/(4\rho_{\gamma})\approx673(1+z)^{-1}$ is the baryon loading with $\rho_b$ and $\rho_\gamma$ being the baryon and photon energy densities, respectively. We remark that, for a given $k$-mode, energy release happens at $k\simeq k_{D}^{}(z)$, where $k_{D}^{}(z)\approx4\times10^{-6}(1+z)^{3/2}\,\mathrm{Mpc}^{-1}$.

Apart from the Silk damping contribution to the effective $\mu$-distortion, another relatively smaller contribution arises from the adiabatic cooling of ordinary matter which continuously extracts energy from the photon bath via Compton scattering in order to establish an equilibrium photon temperature. This leads to a negative $\mu$-distortion of $\mathcal{O}(10^{-9})$, where the tiny magnitude of this energy extraction from the CMB ($\text{d}Q/\text{d}z\vert_{\mathrm{cool}}$) is due to the fact that the heat capacity of the CMB is much larger than that of matter. We follow Refs. \cite{1968ApJ...153....1P,1969JETP...28..146Z,Chluba:2011hw,Chluba:2016bvg} for the computation of this effective energy extraction history.

Several methods have been implemented for the computation of the $\mu$-distortion parameter \cite{Chluba:2016bvg,Mifsud:PBH_PRD}, and we therefore briefly describe our methodology. Given the energy release history\footnote{We neglect the heating rate from tensor perturbations \cite{Chluba:2014qia}, subleading with respect to the considered $\mu$-distortion contributions.} composed of the damping of primordial small-scale perturbations along with the adiabatic cooling of ordinary matter, we can compute the spectral distortion via \cite{Zeldovich:1969ff,Sunyaev:1970er,Chluba:2013pya,Khatri:2013dha}
\begin{equation}
\label{eqn:mu}
\mu=1.4\,\frac{\Delta\rho_{\gamma}}{\rho_{\gamma}}\Bigg\vert_{\mu}=1.4\int_{z_{\mu}}^{z_{\mathrm{max}}}\mathcal{J}_{\mu}\left(z^{\prime}\right)\frac{\mathrm{d}Q}{\mathrm{d}z^{\prime}}\,\mathrm{d}z^{\prime}\;,
\end{equation}
where $\Delta\rho_\gamma/\rho_\gamma\vert_{\mu}$ denotes the effective energy release in the $\mu$-era, and $\mathrm{d}Q/\mathrm{d}z$ is the sum of $\mathrm{d}Q/\mathrm{d}z\vert_{\mathrm{ac}}$ and $\mathrm{d}Q/\mathrm{d}z\vert_{\mathrm{cool}}$. We take the upper integration limit of Eq. (\ref{eqn:mu}) sufficiently behind the thermalization redshift $z_{th}$, in order to take into account that the thermalization efficiency does not abruptly vanish at the thermalization epoch. The transition redshift between the $y$-distortion epoch and the $\mu$-distortion epoch is set by defining the transition redshift $z_\mu$ in Eq. (\ref{eqn:mu}). As described in Ref. \cite{Khatri:2012tw}, a (nearly) pure $\mu$-distortion is created at $y_{\gamma}\gtrsim y_{\gamma}^{\mathrm{max}}=2$, corresponding to $z_{\mu}\approx2\times10^{5}$, where the Compton parameter is defined by
\begin{equation}
\label{eqn:y_gamma}
y_{\gamma}(z)=\int_{0}^{z}\mathrm{d}z^{\prime}\frac{k_{B}^{}\sigma_{T}^{}}{m_{e}c}\frac{n_{e}T_{\gamma}}{H(1+z^{\prime})}\;,
\end{equation}
in which the electron mass is denoted by $m_{e}$. In what follows, we will compute the $\mu$-distortion amplitude by setting the transition redshift in Eq. (\ref{eqn:mu}) equal to the inferred redshift from Eq. (\ref{eqn:y_gamma}), such that $y_\gamma(z_\mu)\approx y_\gamma^\mathrm{max}$. 

Finally, we adopt the distortion visibility function $\mathcal{J}_{\mu}=e^{-\tau(z)}$ of Ref. \cite{Khatri:2012tv}, where $\tau(z)$ is the effective blackbody optical depth. This ensures that the small $\mu$-distortion contribution produced at $z\gtrsim z_{th}$ is also taken into account.

Fig. \ref{fig:mu_window} illustrates the scales probed by various experiments including the $\mu$-distortion measurements, where we plot the normalized dimensionless primordial power spectrum (\ref{eqn:dimensionles_P_zeta}). We here use the $k$-space window function $W_{\mu}(k)\approx2.3\,e^{-2k^2/k_D^2}$ \cite{Pajer:2012vz,Emami:2015xqa}, which accounts for the thermalization process.


\textit{Formation of PBHs \& summary of constraints.}-- Interest in PBHs as dark matter candidates \cite{1975Natur.253..251C,1974MNRAS.168..399C,Bird:2016dcv} has flourished after the first direct detection of gravitational-waves \cite{Abbott:2016blz}, since the source's black hole masses were found to be inconsistent with the typical binary black hole masses that are created from Population I/II main sequence stars \cite{2010ApJ...714.1217B,Prestwich:2007mj,Silverman:2008ss,TheLIGOScientific:2016htt}. After the discovery of similar binary black hole mergers \cite{Abbott:2016nmj,Abbott:2017oio,Abbott:2017gyy,Abbott:2017vtc,LIGOScientific:2018mvr}, the plausible detection of PBHs was gaining ground.

The mechanism behind the formation of PBHs is most likely to be the same phenomenon that shaped the large-scale cosmic structure -- one possibility is indeed from the gravitational collapse of significantly large density fluctuations $(\delta\rho/\rho\sim\mathcal{O}(1))$ that re-enter the horizon during the radiation-dominated era and cannot be overcome by the pressure forces. It is well-known that with a blue spectrum one can produce PBHs, since in this case there is an increase in power on the smallest-scales, particularly on the length scales relevant for PBH formation \cite{Carr:1994ar,Sendouda:2006nu}. In our case we will be using the runnings of the scalar spectral index (refer to Eq. (\ref{eqn:P_zeta})) so that the amplitude of the fluctuations is allowed to increase on the small-scales. As shown in Fig. \ref{fig:mu_window}, the inclusion of the current PBH constraints in our joint data sets would enable us to confront cosmic inflation across a very wide range of length scales.

We now briefly discuss our implemented procedure which incorporates the current PBH constraints, as depicted in Fig. \ref{fig:mu_window}. Provided that at horizon crossing (i.e. $R=(aH)^{-1}$, with $a$ being the scale factor), the relative mass excess inside an overdense region with smoothed density contrast $\delta_{\mathrm{hor}}^{}(R)$, is greater than a critical threshold $\delta_{c}\sim1/3$ \cite{1975ApJ...201....1C}, the region will collapse to form a PBH. We will be using $\delta_c=1/3$ (see Refs. \cite{Polnarev:2006aa,Musco:2004ak} for further details), and we introduce an upper limit for the density contrast of $\delta_{\mathrm{hor}}^{}(R)\lesssim1$ that arises from the possibility of very large density perturbations to close up upon themselves and form separate Universes \cite{1971MNRAS.152...75H,1974MNRAS.168..399C,1975ApJ...201....1C,Harada:2004pe} (see also Ref. \cite{Kopp:2010sh}), although the choice of this upper limit does not alter the abundance of PBHs due to the rapidly decreasing integrands above $\delta_c$. We also consider Gaussian perturbations with the probability distribution of the smoothed density contrast $P_\delta^{}(\delta_{\mathrm{hor}}^{}(R))$, which is given by
\begin{equation}
P_\delta^{}\left(\delta_{\mathrm{hor}^{}}(R)\right)=\frac{1}{\sqrt{2\pi}\sigma_{\delta}^{}(R)}\mathrm{exp}\left(-\frac{\delta^2_{\mathrm{hor}}(R)}{2\sigma^2_{\delta}(R)}\right)\;,
\end{equation}
where the mass variance of the above probability distribution function is given by
\begin{equation}
\label{eqn:mass_variance}
\sigma^2_\delta(R)=\int_0^\infty W^2(kR)\mathcal{P}_\delta^{}(k)\frac{\mathrm{d}k}{k}\;,
\end{equation}
with $W(kR)$ being the Fourier transform of the window function that is used to smooth the density contrast, and $\mathcal{P}_\delta^{}(k)$ denotes the power spectrum of the density contrast. In this work we will be using a Gaussian window function (see Refs. \cite{Blais:2002gw,Bringmann:2001yp} for other alternatives) specified by $W(kR)=\mathrm{exp}\left(-k^2R^2/2\right)$.

We then use the relationship between the power spectra of the density contrast and that of the primordial curvature perturbation $\mathcal{P_R^{}}(k)$, given by \cite{Josan:2009qn}
\begin{equation}
\label{eqn:delta_R}
\mathcal{P}_\delta^{}(k)=\frac{16}{3}\left(\frac{k}{aH}\right)^2j_1^2\left(\frac{k}{\sqrt{3}aH}\right)\mathcal{P_R^{}}(k)\;,
\end{equation}
where $j_1(X)$ is the spherical Bessel function of the first kind. We note that the last quantity in Eq. (\ref{eqn:delta_R}) is identical to the dimensionless power spectrum of Eq. (\ref{eqn:dimensionles_P_zeta}). However, since the integral of the mass variance of Eq. (\ref{eqn:mass_variance}) is dominated by the scales of $k\sim 1/R$, we will assume that over this restricted range of local $k$-values being probed by a specific PBH abundance constraint, the primordial curvature perturbation power spectrum is assumed to be given by a power-law \cite{Drees:2011yz,Drees:2011hb}
\begin{equation}
\label{eqn:PRk}
\mathcal{P_R^{}}(k)=\mathcal{P_R^{}}(k_R^{})\left(\frac{k}{k_R^{}}\right)^{n_s(R)-1}\;,
\end{equation}
with $k_R^{}=1/R$ and
\begin{equation}
\label{eqn:PRkR}
\mathcal{P_R^{}}(k_R^{})=A_s\left(\frac{k_R^{}}{k_0^{}}\right)^{n(R)-1}\;.
\end{equation}
The effective spectral indices $n_s(R)$ and $n(R)$ describe the slope of the power spectrum at the local scales of $k\sim k_R^{}$, and the normalization of the spectrum at $k_R^{}\gg k_0^{}$, respectively. These are related to the primordial curvature power spectrum parameters defined at the pivot scale $k_0^{}$, as follows
\begin{widetext}
\begin{eqnarray}
n(R)&=&n_s-\frac{1}{2}\alpha_s\ln\left(k_{0}^{}R\right)+\frac{1}{6}\beta_s\left[\ln\left(k_{0}^{}R\right)\right]^{2}-\frac{1}{24}\gamma_s\left[\ln\left(k_{0}^{}R\right)\right]^{3}\;\;\;\,,\label{eqn:n_R_PBH}\\[8pt]
n_s(R)&=&n(R)-\frac{1}{2}\alpha_s\ln\left(k_{0}^{}R\right)+\frac{1}{3}\beta_s\left[\ln\left(k_{0}^{}R\right)\right]^{2}-\frac{1}{8}\gamma_s\left[\ln\left(k_{0}^{}R\right)\right]^{3}\,\,,\label{eqn:nsR}
\end{eqnarray}
\end{widetext}
where Eq. (\ref{eqn:n_R_PBH}) is derived by comparing Eq. (\ref{eqn:PRkR}) with Eq. (\ref{eqn:P_zeta}), whereas for the definition of $n_s(R)$ we used the derived expression of $n(R)$ along with the condition of $\mathrm{d}\ln\mathcal{P_R^{}}(k_R^{})/\mathrm{d}\ln k_R^{}=\mathrm{d}\ln \mathcal{P_R^{}}(k)/\mathrm{d}\ln k\vert_{k\,=\,k_R^{}}$. This approach significantly optimized our computations, without altering our results. The final link in the chain is the relationship between the initial PBH mass fraction\footnote{We remark that the critical energy density is denoted by $\rho_\mathrm{cr}$.} $\beta\left(M_{\mathrm{PBH}^{}}\right)=\rho_\mathrm{PBH}^{}(t_i)/\rho_\mathrm{cr}(t_i)$, at the time of PBH formation $t_i$, and the mass variance. In the Press-Schechter formalism \cite{Press:1973iz}, this relationship between the PBH initial mass fraction and the mass variance, is given by
\begin{equation}
\label{eqn:beta_PBH}
\begin{split}
\beta\left(M_{\mathrm{PBH}}\right)&=2\frac{M_{\mathrm{PBH}}}{M_{\mathrm{H}}}\int_{\delta_c}^{1}P_\delta^{}\left(\delta_{\mathrm{hor}^{}}(R)\right)\,\mathrm{d}\delta_{\mathrm{hor}}(R)\\
&\approx\gamma\,\mathrm{erfc}\left(\frac{\delta_c}{\sqrt{2}\sigma_\delta^{}(R)}\right)\;,
\end{split}
\end{equation}
where $\mathrm{erfc}(X)$ is the complementary error function. In Eq. (\ref{eqn:beta_PBH}) we adopted the assumption that the PBHs form at a single epoch and that their mass is a fixed fraction $\gamma$, of the horizon mass $M_\mathrm{H}^{}=(4\pi/3)\rho H^{-3}$, such that $M_{\mathrm{PBH}}=\gamma\,M_{\mathrm{H}}^{}$. The Friedmann equation in a radiation-dominated era reduces to $H^2\approx(8\pi G/3)\rho$, with $\rho$ denoting the total radiation energy density and $G$ is Newton's gravitational constant, leading to the approximate relationship of $M_{\mathrm{PBH}}\approx1.97\times10^5\,\gamma\,(t/1\,\mathrm{s})M_\odot$. The latter relationship clearly shows that PBHs span a huge mass range, that is determined by their time of formation, which features in the PBH abundance constraints. Moreover, we make use of $\gamma\approx3^{-3/2}$ \cite{1975ApJ...201....1C}, which is derived from simple analytical calculations, although it depends on the details of the gravitational collapse mechanism.

\begin{figure}[t]
\begin{center}
\includegraphics[width=0.99\columnwidth]{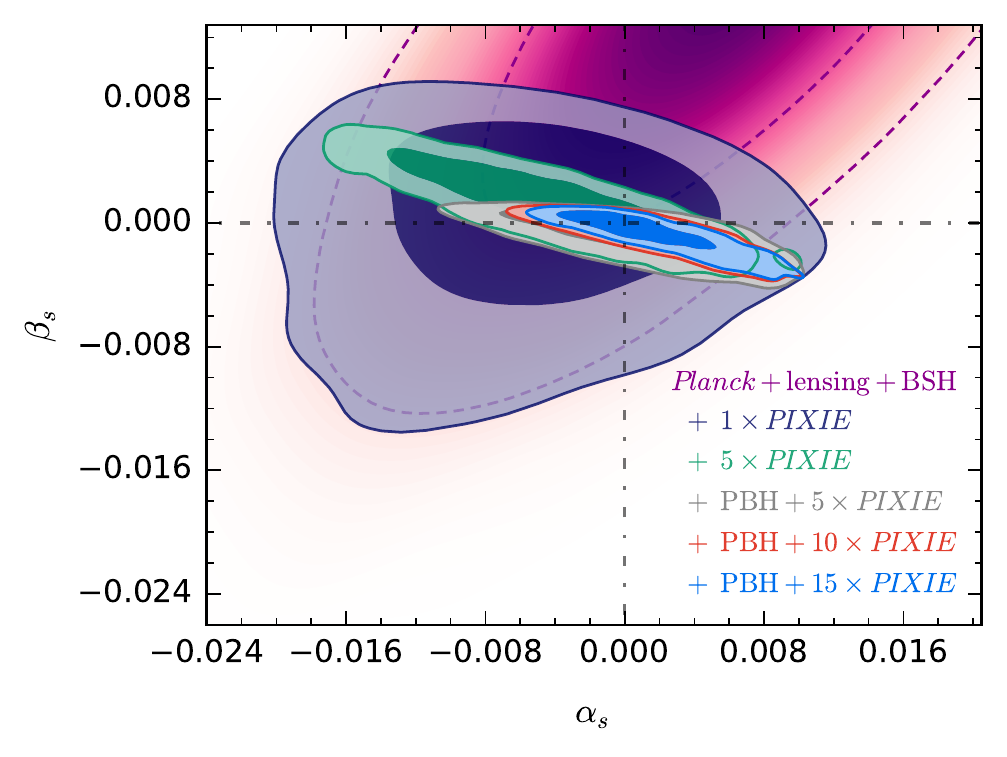}
\caption{We show the tightest constraints in the $\alpha_s$--$\beta_s$ plane inferred from the joint data sets indicated in the figure. The dot-dashed lines illustrate the null runnings of the scalar spectral index assumed in the $\Lambda$CDM model.
\label{fig:alpha_beta_comp}}
\end{center}
\end{figure}

Finally, the above Friedmann equation in the radiation-dominated epoch along with cosmic expansion at constant entropy $(\rho\propto g_\ast^{-1/3}a^{-4}$ \cite{Green:2004wb}, with $g_\ast$ denoting the number of relativistic degrees of freedom), imply that
\begin{equation}
M_\mathrm{PBH}=\gamma M_\mathrm{H,eq}\left(\frac{g_{\ast,\mathrm{eq}}}{g_\ast}\right)^{1/3}\left(k_\mathrm{eq}R\right)^2\;,
\end{equation}
where $M_\mathrm{H,eq}$ is the horizon mass at matter-radiation equality, the comoving wavenumber at equality is denoted by $k_{\mathrm{eq}}$, and $a_{\mathrm{eq}}$ is the cosmic scale factor at equality. Thus, the crucial relationship between the PBH mass and the comoving smoothing scale $R$, is given by
\begin{equation}
\begin{split}
\frac{R}{1\,\mathrm{Mpc}}\approx\hspace{2pt}&3.70\times10^{-23}\gamma^{-1/2}\left(\Omega_{r}h^2\right)^{-1/2}a_{\mathrm{eq}}^{1/2}\\
&\times\left(\frac{k_{\mathrm{eq}}}{1\,\mathrm{Mpc}^{-1}}\right)^{1/2}\left(\frac{M_{\mathrm{PBH}}}{1\,\mathrm{g}}\right)^{1/2}\;,
\end{split}
\end{equation}
where we used $g_{\ast,\mathrm{eq}}\approx3.36$ as the number of relativistic degrees of freedom at the time of matter-radiation equality.

{\setlength\extrarowheight{8pt}
\begin{table*}[t!]
\begin{center}
\begin{tabular}{ l @{\hspace{0.5\tabcolsep}} c @{\hspace{0.5\tabcolsep}}  c @{\hspace{0.5\tabcolsep}} c @{\hspace{0.5\tabcolsep}} c @{\hspace{0.5\tabcolsep}} c }
\hline
\hline
\multirow{2}{*}{Parameter}	& ~\begin{tabular}[t]{@{}c@{}}\textit{Planck} + lensing \\ + BSH + PBH\end{tabular}~ 
							& ~\multirow{2}{*}{+ $1\,\times$\textit{PIXIE}}~ 
							& ~\multirow{2}{*}{+ $5\,\times$\textit{PIXIE}}~
							& ~\multirow{2}{*}{+ $10\,\times$\textit{PIXIE}}~
							& ~\multirow{2}{*}{+ $15\,\times$\textit{PIXIE}}~ \\ 
\hline
$\ln\left(10^{10}A_s\right)$\dotfill & $3.0531^{+0.0233}_{-0.0242}$ & $3.0528^{+0.0228}_{-0.0223}$ & $3.0515^{+0.0233}_{-0.0228}$ & $3.0510^{+0.0240}_{-0.0232}$ & $3.0509^{+0.0244}_{-0.0235}$ \\

$n_s$\dotfill & $0.96320^{+0.00442}_{-0.00446}$ & $0.96321^{+0.00415}_{-0.00422}$ & $0.96386^{+0.00393}_{-0.00387}$ & $0.96411^{+0.00387}_{-0.00379}$ & $0.96418^{+0.00384}_{-0.00375}$ \\ 

$\alpha_s$\dotfill & $-0.00687^{+0.00776}_{-0.00737}$~~ & $-0.00537^{+0.00680}_{-0.00668}$~~ & $0.00023^{+0.00398}_{-0.00439}$ & $0.00113^{+0.00266}_{-0.00402}$ & $0.00131^{+0.00217}_{-0.00397}$ \\ 

$\beta_s$\dotfill & $-0.00496^{+0.00686}_{-0.00153}$~~ & $-0.00364^{+0.00451}_{-0.00108}$~~ & $-0.00069^{+0.00142}_{-0.00052}$~~ & $-0.00059^{+0.00125}_{-0.00045}$~~ & $-0.00058^{+0.00122}_{-0.00042}$~~ \\[1.ex]
\hline
\hline
\end{tabular}
\end{center}
\caption{\label{table:primordial_alpha_beta_IS} We report the 68\% CLs for the $\Lambda$CDM + $\alpha_s$ + $\beta_s$ model. The inferred constraints from $n\times$\textit{PIXIE} are obtained by post-processing the Markov chains with a Gaussian likelihood of $\mu=(1.48\pm1.00/n)\times10^{-8}$, where its mean value is fixed to the derived mean value of the $\mu$-type spectral distortion parameter in the $\Lambda$CDM model.}
\end{table*}}

From the full compilation of the $\beta\left(M_\mathrm{PBH}\right)$ constraints presented in Fig. \ref{fig:mu_window}, we derive an upper bound on $\sigma_\delta^{}(R)$ by inverting Eq. (\ref{eqn:beta_PBH}). We then compute the mass variance for the given set of cosmological parameters using Eqs. (\ref{eqn:mass_variance})--(\ref{eqn:nsR}), and check if the inferred scale-dependent upper bound on the mass variance is satisfied over all PBH mass scales. Further details regarding these PBH constraints which are all associated with various caveats \cite{Carr:2009jm,Barnacka:2012bm,Graham:2015apa,Capela:2013yf,Niikura:2017zjd,Griest:2013aaa,Tisserand:2006zx,Oguri:2017ock,Allsman:2000kg,Quinn:2009zg,Monroy-Rodriguez:2014ula,Koushiappas:2017chw,Ali-Haimoud:2016mbv,Brandt:2016aco,Inoue:2017csr,Wilkinson:2001vv,Carr:2016drx,Ade:2015xua,Carr:1994ar,Khlopov:2015oda,Lehoucq:2009ge,Carr:2016hva,Mack:2008nv}, can be found in the companion article \cite{Mifsud:PBH_PRD}.

\textit{Implications for cosmic inflation.}-- We now infer the posterior distributions and the confidence limits (CLs) on the primordial power spectrum parameters of Eq. (\ref{eqn:P_zeta}). We further vary $\Omega_bh^2$, the cold dark matter density parameter $\Omega_ch^2$, the ratio of the sound horizon to the angular diameter distance at decoupling $\theta_s$, and the reionization optical depth $\tau_{\mathrm{reio}}$, for which we specify flat priors.

We created the Markov chain Monte Carlo (MCMC) samples using a customized version of \texttt{CLASS} \cite{Blas:2011rf} along with \texttt{Monte Python} \cite{Audren:2012wb}, and then fully analysed the chains with \texttt{GetDist} \cite{Lewis:2002ah}. Moreover, we inferred the $\mu$-distortion via importance sampling of the MCMC samples implemented in the \texttt{GetDist} routine, where we accurately calculated the free electron fraction $X_e$, by interfacing the modified \texttt{IDISTORT} code \cite{Khatri:2012tw} with the primordial recombination \texttt{HyRec} code \cite{PhysRevD.83.043513}.

Further to the PBH constraint of Fig. \ref{fig:mu_window}, which was implemented via a step-function likelihood, we will be using the \textit{Planck} 2015 temperature and polarization (TT, TE, EE) high-$\ell$ and low-$\ell$ likelihoods \cite{Aghanim:2015xee}, along with the lensing likelihood \cite{Ade:2015zua}. We will refer to the \textit{Planck} joint likelihoods by \textit{Planck} + lensing. Moreover, we will further consider a background data set which we denote by BSH. This consists of baryon acoustic oscillation measurements \cite{Beutler:2011hx,Ross:2014qpa,Ata:2017dya,Bautista:2017zgn,Alam:2016hwk}, a supernovae Type Ia sample \cite{Betoule:2014frx}, and a cosmic chronometers data set \cite{Simon:2004tf,Stern:2009ep,Zhang:2012mp,Moresco:2012jh,Moresco:2015cya,Moresco:2016mzx}.

In order to study the implications of prospective measurements of the $\mu$-distortion, we make our forecast around the $\Lambda$CDM prediction \cite{Cabass:2016giw,Cabass:2016ldu}, in which we post-process the Markov chains with a Gaussian likelihood considering $\mu_\mathrm{fid}^{}=(1.48\pm1.00/n)\times10^{-8}$ for an $n\times$\textit{PIXIE} sensitivity, where we set the fiducial mean value to coincide with the derived mean-fit value in the $\Lambda$CDM scenario. We have set the $1\sigma$ error of $\mu_\mathrm{fid}^{}$ to be in agreement with the reported $1\times$\textit{PIXIE}-type experiment uncertainty of Refs. \cite{Kogut:2011xw,Chluba:2013pya}, whereas a $10\times$\textit{PIXIE} spectral sensitivity could possibly be reached by a \textit{PRISM}-like \cite{Andre:2013afa} experiment. We also examine a very optimistic $15\times$\textit{PIXIE}-type experiment in order to discuss its quantitative improvement on our constraints.

In the $\Lambda$CDM + $\alpha_s$ + $\beta_s$ model, we find that the large-scale data sets favour positive $\alpha_s$ and $\beta_s$, as depicted by the dashed confidence region of Fig. \ref{fig:alpha_beta_comp}. In the case of $\alpha_s$, the derived constraint is in a very good agreement with a null running of the scalar spectral index $(<1\sigma)$, however $\beta_s$ is found to be greater than zero at $\sim1.4$ standard deviations. Thus, in this extended model, it is evident that these positive values of $\alpha_s$ and $\beta_s$ will be able to significantly increase the power on the small-scales. As a result, the PBH constraint is crucial in this case since the PBH abundance constraint, although relatively weak, will not allow for these positive values of the runnings of the scalar spectral index. Indeed, when we confront the $\Lambda$CDM + $\alpha_s$ + $\beta_s$ model with the \textit{Planck} + lensing + BSH + PBH joint likelihood, the PBH upper bound is able to push the constraints on both $\alpha_s$ and $\beta_s$ to negative values, which are found to be consistent with zero at less than one standard deviation.

In Table \ref{table:primordial_alpha_beta_IS} we present the 68\% CLs for this model, and in Fig. \ref{fig:alpha_beta_comp} we illustrate the remarkable shrinking of the marginalized contours when we consider the large-scale data sets along with the crucial information from the small-scales. The 68\% CLs on the running-of-the-running of the scalar spectral index are $-5\times10^{-3}\lesssim\beta_s\lesssim9\times10^{-4}$ (\textit{Planck} + lensing + BSH + PBH + $1\times$\textit{PIXIE}), or $-1.2\times10^{-3}\lesssim\beta_s\lesssim7.3\times10^{-4}$ (\textit{Planck} + lensing + BSH + PBH + $5\times$\textit{PIXIE}), or $-1.0\times10^{-3}\lesssim\beta_s\lesssim6.4\times10^{-4}$ (\textit{Planck} + lensing + BSH + PBH + $15\times$\textit{PIXIE}), implying that the small-scale data can robustly constrain higher-order parameters of the primordial curvature power spectrum. 


\textit{Conclusion.}-- By combining the CMB anisotropies data with the small-scale constraints from the abundance of PBHs along with prospective $\mu$-distortion measurements, we placed tight limits on the higher-order parameters of the primordial power spectrum. We showed that these independent probes are able to exclude a significantly large portion of the currently viable region in the $\alpha_s$--$\beta_s$ plane that is solely inferred from large-scale experiments, possibly ruling out a number of cosmic inflationary models. Thus, a better understanding of the physics of PBHs along with future surveys that would be able to probe the thermal history of our early Universe and the primordial gravitational-wave spectrum across several decades in frequency \cite{Lasky:2015lej,Meerburg:2015zua,Cabass:2015jwe,Wang:2016tbj,Guzzetti:2016mkm,Caprini:2018mtu}, are paramount for constraining the plethora of governing theories of the early Universe.

\begin{acknowledgments}
The work of CvdB is supported by the Lancaster-Manchester-Sheffield Consortium for Fundamental Physics under STFC Grant No. ST/L000520/1.
\end{acknowledgments}

\bibliography{fullbib}

\begin{thebibliography}{118}%
\makeatletter
\providecommand \@ifxundefined [1]{%
 \@ifx{#1\undefined}
}%
\providecommand \@ifnum [1]{%
 \ifnum #1\expandafter \@firstoftwo
 \else \expandafter \@secondoftwo
 \fi
}%
\providecommand \@ifx [1]{%
 \ifx #1\expandafter \@firstoftwo
 \else \expandafter \@secondoftwo
 \fi
}%
\providecommand \natexlab [1]{#1}%
\providecommand \enquote  [1]{``#1''}%
\providecommand \bibnamefont  [1]{#1}%
\providecommand \bibfnamefont [1]{#1}%
\providecommand \citenamefont [1]{#1}%
\providecommand \href@noop [0]{\@secondoftwo}%
\providecommand \href [0]{\begingroup \@sanitize@url \@href}%
\providecommand \@href[1]{\@@startlink{#1}\@@href}%
\providecommand \@@href[1]{\endgroup#1\@@endlink}%
\providecommand \@sanitize@url [0]{\catcode `\\12\catcode `\$12\catcode
  `\&12\catcode `\#12\catcode `\^12\catcode `\_12\catcode `\%12\relax}%
\providecommand \@@startlink[1]{}%
\providecommand \@@endlink[0]{}%
\providecommand \url  [0]{\begingroup\@sanitize@url \@url }%
\providecommand \@url [1]{\endgroup\@href {#1}{\urlprefix }}%
\providecommand \urlprefix  [0]{URL }%
\providecommand \Eprint [0]{\href }%
\providecommand \doibase [0]{http://dx.doi.org/}%
\providecommand \selectlanguage [0]{\@gobble}%
\providecommand \bibinfo  [0]{\@secondoftwo}%
\providecommand \bibfield  [0]{\@secondoftwo}%
\providecommand \translation [1]{[#1]}%
\providecommand \BibitemOpen [0]{}%
\providecommand \bibitemStop [0]{}%
\providecommand \bibitemNoStop [0]{.\EOS\space}%
\providecommand \EOS [0]{\spacefactor3000\relax}%
\providecommand \BibitemShut  [1]{\csname bibitem#1\endcsname}%
\let\auto@bib@innerbib\@empty
\bibitem [{\citenamefont {Ade}\ \emph {et~al.}(2016{\natexlab{a}})\citenamefont
  {Ade} \emph {et~al.}}]{Ade:2015xua}%
  \BibitemOpen
  \bibfield  {author} {\bibinfo {author} {\bibfnamefont {P.~A.~R.}\
  \bibnamefont {Ade}} \emph {et~al.} (\bibinfo {collaboration} {Planck}),\
  }\href {\doibase 10.1051/0004-6361/201525830} {\bibfield  {journal} {\bibinfo
   {journal} {Astron. Astrophys.}\ }\textbf {\bibinfo {volume} {594}},\
  \bibinfo {pages} {A13} (\bibinfo {year} {2016}{\natexlab{a}})},\ \Eprint
  {http://arxiv.org/abs/1502.01589} {arXiv:1502.01589 [astro-ph.CO]}
  \BibitemShut {NoStop}%
\bibitem [{\citenamefont {Starobinsky}(1979)}]{Starobinsky:1979ty}%
  \BibitemOpen
  \bibfield  {author} {\bibinfo {author} {\bibfnamefont {A.~A.}\ \bibnamefont
  {Starobinsky}},\ }\href@noop {} {\bibfield  {journal} {\bibinfo  {journal}
  {JETP Lett.}\ }\textbf {\bibinfo {volume} {30}},\ \bibinfo {pages} {682}
  (\bibinfo {year} {1979})}\BibitemShut {NoStop}%
\bibitem [{\citenamefont {Kazanas}(1980)}]{Kazanas:1980tx}%
  \BibitemOpen
  \bibfield  {author} {\bibinfo {author} {\bibfnamefont {D.}~\bibnamefont
  {Kazanas}},\ }\href {\doibase 10.1086/183361} {\bibfield  {journal} {\bibinfo
   {journal} {Astrophys. J.}\ }\textbf {\bibinfo {volume} {241}},\ \bibinfo
  {pages} {L59} (\bibinfo {year} {1980})}\BibitemShut {NoStop}%
\bibitem [{\citenamefont {Guth}(1981)}]{Guth:1980zm}%
  \BibitemOpen
  \bibfield  {author} {\bibinfo {author} {\bibfnamefont {A.~H.}\ \bibnamefont
  {Guth}},\ }\href {\doibase 10.1103/PhysRevD.23.347} {\bibfield  {journal}
  {\bibinfo  {journal} {Phys. Rev.}\ }\textbf {\bibinfo {volume} {D23}},\
  \bibinfo {pages} {347} (\bibinfo {year} {1981})}\BibitemShut {NoStop}%
\bibitem [{\citenamefont {Linde}(1982)}]{Linde:1981mu}%
  \BibitemOpen
  \bibfield  {author} {\bibinfo {author} {\bibfnamefont {A.~D.}\ \bibnamefont
  {Linde}},\ }\href {\doibase 10.1016/0370-2693(82)91219-9} {\bibfield
  {journal} {\bibinfo  {journal} {Phys. Lett.}\ }\textbf {\bibinfo {volume}
  {108B}},\ \bibinfo {pages} {389} (\bibinfo {year} {1982})}\BibitemShut
  {NoStop}%
\bibitem [{\citenamefont {Sato}(1981)}]{Sato:1980yn}%
  \BibitemOpen
  \bibfield  {author} {\bibinfo {author} {\bibfnamefont {K.}~\bibnamefont
  {Sato}},\ }\href@noop {} {\bibfield  {journal} {\bibinfo  {journal} {Mon.
  Not. Roy. Astron. Soc.}\ }\textbf {\bibinfo {volume} {195}},\ \bibinfo
  {pages} {467} (\bibinfo {year} {1981})}\BibitemShut {NoStop}%
\bibitem [{\citenamefont {Albrecht}\ and\ \citenamefont
  {Steinhardt}(1982)}]{Albrecht:1982wi}%
  \BibitemOpen
  \bibfield  {author} {\bibinfo {author} {\bibfnamefont {A.}~\bibnamefont
  {Albrecht}}\ and\ \bibinfo {author} {\bibfnamefont {P.~J.}\ \bibnamefont
  {Steinhardt}},\ }\href {\doibase 10.1103/PhysRevLett.48.1220} {\bibfield
  {journal} {\bibinfo  {journal} {Phys. Rev. Lett.}\ }\textbf {\bibinfo
  {volume} {48}},\ \bibinfo {pages} {1220} (\bibinfo {year}
  {1982})}\BibitemShut {NoStop}%
\bibitem [{\citenamefont {Martin}\ \emph {et~al.}(2014)\citenamefont {Martin},
  \citenamefont {Ringeval}, \citenamefont {Trotta},\ and\ \citenamefont
  {Vennin}}]{Martin:2013nzq}%
  \BibitemOpen
  \bibfield  {author} {\bibinfo {author} {\bibfnamefont {J.}~\bibnamefont
  {Martin}}, \bibinfo {author} {\bibfnamefont {C.}~\bibnamefont {Ringeval}},
  \bibinfo {author} {\bibfnamefont {R.}~\bibnamefont {Trotta}}, \ and\ \bibinfo
  {author} {\bibfnamefont {V.}~\bibnamefont {Vennin}},\ }\href {\doibase
  10.1088/1475-7516/2014/03/039} {\bibfield  {journal} {\bibinfo  {journal}
  {JCAP}\ }\textbf {\bibinfo {volume} {1403}},\ \bibinfo {pages} {039}
  (\bibinfo {year} {2014})},\ \Eprint {http://arxiv.org/abs/1312.3529}
  {arXiv:1312.3529 [astro-ph.CO]} \BibitemShut {NoStop}%
\bibitem [{\citenamefont {Guth}\ \emph {et~al.}(2014)\citenamefont {Guth},
  \citenamefont {Kaiser},\ and\ \citenamefont {Nomura}}]{Guth:2013sya}%
  \BibitemOpen
  \bibfield  {author} {\bibinfo {author} {\bibfnamefont {A.~H.}\ \bibnamefont
  {Guth}}, \bibinfo {author} {\bibfnamefont {D.~I.}\ \bibnamefont {Kaiser}}, \
  and\ \bibinfo {author} {\bibfnamefont {Y.}~\bibnamefont {Nomura}},\ }\href
  {\doibase 10.1016/j.physletb.2014.03.020} {\bibfield  {journal} {\bibinfo
  {journal} {Phys. Lett.}\ }\textbf {\bibinfo {volume} {B733}},\ \bibinfo
  {pages} {112} (\bibinfo {year} {2014})},\ \Eprint
  {http://arxiv.org/abs/1312.7619} {arXiv:1312.7619 [astro-ph.CO]} \BibitemShut
  {NoStop}%
\bibitem [{\citenamefont {Burgess}\ \emph {et~al.}(2013)\citenamefont
  {Burgess}, \citenamefont {Cicoli},\ and\ \citenamefont
  {Quevedo}}]{Burgess:2013sla}%
  \BibitemOpen
  \bibfield  {author} {\bibinfo {author} {\bibfnamefont {C.~P.}\ \bibnamefont
  {Burgess}}, \bibinfo {author} {\bibfnamefont {M.}~\bibnamefont {Cicoli}}, \
  and\ \bibinfo {author} {\bibfnamefont {F.}~\bibnamefont {Quevedo}},\ }\href
  {\doibase 10.1088/1475-7516/2013/11/003} {\bibfield  {journal} {\bibinfo
  {journal} {JCAP}\ }\textbf {\bibinfo {volume} {1311}},\ \bibinfo {pages}
  {003} (\bibinfo {year} {2013})},\ \Eprint {http://arxiv.org/abs/1306.3512}
  {arXiv:1306.3512 [hep-th]} \BibitemShut {NoStop}%
\bibitem [{\citenamefont {Akrami}\ \emph {et~al.}(2018)\citenamefont {Akrami}
  \emph {et~al.}}]{Akrami:2018odb}%
  \BibitemOpen
  \bibfield  {author} {\bibinfo {author} {\bibfnamefont {Y.}~\bibnamefont
  {Akrami}} \emph {et~al.} (\bibinfo {collaboration} {Planck}),\ }\href@noop {}
  {\bibfield  {journal} {\bibinfo  {journal} {ArXiv e-prints}\ } (\bibinfo
  {year} {2018})},\ \Eprint {http://arxiv.org/abs/1807.06211} {arXiv:1807.06211
  [astro-ph.CO]} \BibitemShut {NoStop}%
\bibitem [{\citenamefont {Weymann}(1966)}]{Weymann1966}%
  \BibitemOpen
  \bibfield  {author} {\bibinfo {author} {\bibfnamefont {R.}~\bibnamefont
  {Weymann}},\ }\href@noop {} {\bibfield  {journal} {\bibinfo  {journal}
  {Astrophys. J.}\ }\textbf {\bibinfo {volume} {145}},\ \bibinfo {pages} {560}
  (\bibinfo {year} {1966})}\BibitemShut {NoStop}%
\bibitem [{\citenamefont {Zel'dovich}\ and\ \citenamefont
  {Sunyaev}(1969)}]{Zeldovich:1969ff}%
  \BibitemOpen
  \bibfield  {author} {\bibinfo {author} {\bibfnamefont {Y.~B.}\ \bibnamefont
  {Zel'dovich}}\ and\ \bibinfo {author} {\bibfnamefont {R.~A.}\ \bibnamefont
  {Sunyaev}},\ }\href {\doibase 10.1007/BF00661821} {\bibfield  {journal}
  {\bibinfo  {journal} {Astrophys. Space Sci.}\ }\textbf {\bibinfo {volume}
  {4}},\ \bibinfo {pages} {301} (\bibinfo {year} {1969})}\BibitemShut {NoStop}%
\bibitem [{\citenamefont {Sunyaev}\ and\ \citenamefont
  {Zel'dovich}(1970)}]{Sunyaev:1970er}%
  \BibitemOpen
  \bibfield  {author} {\bibinfo {author} {\bibfnamefont {R.~A.}\ \bibnamefont
  {Sunyaev}}\ and\ \bibinfo {author} {\bibfnamefont {Y.~B.}\ \bibnamefont
  {Zel'dovich}},\ }\href@noop {} {\bibfield  {journal} {\bibinfo  {journal}
  {Astrophys. Space Sci.}\ }\textbf {\bibinfo {volume} {7}},\ \bibinfo {pages}
  {20} (\bibinfo {year} {1970})}\BibitemShut {NoStop}%
\bibitem [{\citenamefont {Daly}(1991)}]{Daly1991}%
  \BibitemOpen
  \bibfield  {author} {\bibinfo {author} {\bibfnamefont {R.~A.}\ \bibnamefont
  {Daly}},\ }\href@noop {} {\bibfield  {journal} {\bibinfo  {journal}
  {Astrophys. J.}\ }\textbf {\bibinfo {volume} {371}},\ \bibinfo {pages} {14}
  (\bibinfo {year} {1991})}\BibitemShut {NoStop}%
\bibitem [{\citenamefont {Barrow}\ and\ \citenamefont
  {Coles}(1991)}]{Barrow_Coles1991}%
  \BibitemOpen
  \bibfield  {author} {\bibinfo {author} {\bibfnamefont {J.~D.}\ \bibnamefont
  {Barrow}}\ and\ \bibinfo {author} {\bibfnamefont {P.}~\bibnamefont {Coles}},\
  }\href@noop {} {\bibfield  {journal} {\bibinfo  {journal} {Mon. Not. Roy.
  Astron. Soc.}\ }\textbf {\bibinfo {volume} {248}},\ \bibinfo {pages} {52}
  (\bibinfo {year} {1991})}\BibitemShut {NoStop}%
\bibitem [{\citenamefont {Zel'dovich}\ and\ \citenamefont
  {Novikov}(1967)}]{1967SvA....10..602Z}%
  \BibitemOpen
  \bibfield  {author} {\bibinfo {author} {\bibfnamefont {Y.~B.}\ \bibnamefont
  {Zel'dovich}}\ and\ \bibinfo {author} {\bibfnamefont {I.~D.}\ \bibnamefont
  {Novikov}},\ }\href@noop {} {\bibfield  {journal} {\bibinfo  {journal}
  {Soviet Ast.}\ }\textbf {\bibinfo {volume} {10}},\ \bibinfo {pages} {{602}}
  (\bibinfo {year} {1967})}\BibitemShut {NoStop}%
\bibitem [{\citenamefont {Hawking}(1971)}]{1971MNRAS.152...75H}%
  \BibitemOpen
  \bibfield  {author} {\bibinfo {author} {\bibfnamefont {S.~W.}\ \bibnamefont
  {Hawking}},\ }\href@noop {} {\bibfield  {journal} {\bibinfo  {journal} {Mon.
  Not. Roy. Astron. Soc.}\ }\textbf {\bibinfo {volume} {152}},\ \bibinfo
  {pages} {75} (\bibinfo {year} {1971})}\BibitemShut {NoStop}%
\bibitem [{\citenamefont {Hawking}(1974)}]{1974Natur.248...30H}%
  \BibitemOpen
  \bibfield  {author} {\bibinfo {author} {\bibfnamefont {S.~W.}\ \bibnamefont
  {Hawking}},\ }\href@noop {} {\bibfield  {journal} {\bibinfo  {journal}
  {Nature}\ }\textbf {\bibinfo {volume} {248}},\ \bibinfo {pages} {30}
  (\bibinfo {year} {1974})}\BibitemShut {NoStop}%
\bibitem [{\citenamefont {Carr}\ and\ \citenamefont
  {Hawking}(1974)}]{1974MNRAS.168..399C}%
  \BibitemOpen
  \bibfield  {author} {\bibinfo {author} {\bibfnamefont {B.~J.}\ \bibnamefont
  {Carr}}\ and\ \bibinfo {author} {\bibfnamefont {S.~W.}\ \bibnamefont
  {Hawking}},\ }\href@noop {} {\bibfield  {journal} {\bibinfo  {journal} {Mon.
  Not. Roy. Astron. Soc.}\ }\textbf {\bibinfo {volume} {168}},\ \bibinfo
  {pages} {399} (\bibinfo {year} {1974})}\BibitemShut {NoStop}%
\bibitem [{\citenamefont {Hu}\ and\ \citenamefont {Silk}(1993)}]{Hu:1993gc}%
  \BibitemOpen
  \bibfield  {author} {\bibinfo {author} {\bibfnamefont {W.}~\bibnamefont
  {Hu}}\ and\ \bibinfo {author} {\bibfnamefont {J.}~\bibnamefont {Silk}},\
  }\href {\doibase 10.1103/PhysRevLett.70.2661} {\bibfield  {journal} {\bibinfo
   {journal} {Phys. Rev. Lett.}\ }\textbf {\bibinfo {volume} {70}},\ \bibinfo
  {pages} {2661} (\bibinfo {year} {1993})}\BibitemShut {NoStop}%
\bibitem [{\citenamefont {Chen}\ and\ \citenamefont
  {Kamionkowski}(2004)}]{Chen:2003gz}%
  \BibitemOpen
  \bibfield  {author} {\bibinfo {author} {\bibfnamefont {X.-L.}\ \bibnamefont
  {Chen}}\ and\ \bibinfo {author} {\bibfnamefont {M.}~\bibnamefont
  {Kamionkowski}},\ }\href {\doibase 10.1103/PhysRevD.70.043502} {\bibfield
  {journal} {\bibinfo  {journal} {Phys. Rev.}\ }\textbf {\bibinfo {volume}
  {D70}},\ \bibinfo {pages} {043502} (\bibinfo {year} {2004})},\ \Eprint
  {http://arxiv.org/abs/astro-ph/0310473} {arXiv:astro-ph/0310473 [astro-ph]}
  \BibitemShut {NoStop}%
\bibitem [{\citenamefont {Padmanabhan}\ and\ \citenamefont
  {Finkbeiner}(2005)}]{Padmanabhan:2005es}%
  \BibitemOpen
  \bibfield  {author} {\bibinfo {author} {\bibfnamefont {N.}~\bibnamefont
  {Padmanabhan}}\ and\ \bibinfo {author} {\bibfnamefont {D.~P.}\ \bibnamefont
  {Finkbeiner}},\ }\href {\doibase 10.1103/PhysRevD.72.023508} {\bibfield
  {journal} {\bibinfo  {journal} {Phys. Rev.}\ }\textbf {\bibinfo {volume}
  {D72}},\ \bibinfo {pages} {023508} (\bibinfo {year} {2005})},\ \Eprint
  {http://arxiv.org/abs/astro-ph/0503486} {arXiv:astro-ph/0503486 [astro-ph]}
  \BibitemShut {NoStop}%
\bibitem [{\citenamefont {Chluba}(2010)}]{Chluba:2009uv}%
  \BibitemOpen
  \bibfield  {author} {\bibinfo {author} {\bibfnamefont {J.}~\bibnamefont
  {Chluba}},\ }\href {\doibase 10.1111/j.1365-2966.2009.15957.x} {\bibfield
  {journal} {\bibinfo  {journal} {Mon. Not. Roy. Astron. Soc.}\ }\textbf
  {\bibinfo {volume} {402}},\ \bibinfo {pages} {1195} (\bibinfo {year}
  {2010})},\ \Eprint {http://arxiv.org/abs/0910.3663} {arXiv:0910.3663
  [astro-ph.CO]} \BibitemShut {NoStop}%
\bibitem [{\citenamefont {Silk}(1968)}]{Silk:1967kq}%
  \BibitemOpen
  \bibfield  {author} {\bibinfo {author} {\bibfnamefont {J.}~\bibnamefont
  {Silk}},\ }\href {\doibase 10.1086/149449} {\bibfield  {journal} {\bibinfo
  {journal} {Astrophys. J.}\ }\textbf {\bibinfo {volume} {151}},\ \bibinfo
  {pages} {459} (\bibinfo {year} {1968})}\BibitemShut {NoStop}%
\bibitem [{\citenamefont {Kogut}\ \emph {et~al.}(2011)\citenamefont {Kogut}
  \emph {et~al.}}]{Kogut:2011xw}%
  \BibitemOpen
  \bibfield  {author} {\bibinfo {author} {\bibfnamefont {A.}~\bibnamefont
  {Kogut}} \emph {et~al.},\ }\href {\doibase 10.1088/1475-7516/2011/07/025}
  {\bibfield  {journal} {\bibinfo  {journal} {JCAP}\ }\textbf {\bibinfo
  {volume} {1107}},\ \bibinfo {pages} {025} (\bibinfo {year} {2011})},\ \Eprint
  {http://arxiv.org/abs/1105.2044} {arXiv:1105.2044 [astro-ph.CO]} \BibitemShut
  {NoStop}%
\bibitem [{\citenamefont {Chluba}\ and\ \citenamefont
  {Sunyaev}(2012)}]{Chluba:2011hw}%
  \BibitemOpen
  \bibfield  {author} {\bibinfo {author} {\bibfnamefont {J.}~\bibnamefont
  {Chluba}}\ and\ \bibinfo {author} {\bibfnamefont {R.~A.}\ \bibnamefont
  {Sunyaev}},\ }\href {\doibase 10.1111/j.1365-2966.2011.19786.x} {\bibfield
  {journal} {\bibinfo  {journal} {Mon. Not. Roy. Astron. Soc.}\ }\textbf
  {\bibinfo {volume} {419}},\ \bibinfo {pages} {1294} (\bibinfo {year}
  {2012})},\ \Eprint {http://arxiv.org/abs/1109.6552} {arXiv:1109.6552
  [astro-ph.CO]} \BibitemShut {NoStop}%
\bibitem [{\citenamefont {Khatri}\ and\ \citenamefont
  {Sunyaev}(2012{\natexlab{a}})}]{Khatri:2012tv}%
  \BibitemOpen
  \bibfield  {author} {\bibinfo {author} {\bibfnamefont {R.}~\bibnamefont
  {Khatri}}\ and\ \bibinfo {author} {\bibfnamefont {R.~A.}\ \bibnamefont
  {Sunyaev}},\ }\href {\doibase 10.1088/1475-7516/2012/06/038} {\bibfield
  {journal} {\bibinfo  {journal} {JCAP}\ }\textbf {\bibinfo {volume} {1206}},\
  \bibinfo {pages} {038} (\bibinfo {year} {2012}{\natexlab{a}})},\ \Eprint
  {http://arxiv.org/abs/1203.2601} {arXiv:1203.2601 [astro-ph.CO]} \BibitemShut
  {NoStop}%
\bibitem [{\citenamefont {Chluba}\ \emph {et~al.}(2012)\citenamefont {Chluba},
  \citenamefont {Khatri},\ and\ \citenamefont {Sunyaev}}]{Chluba:2012gq}%
  \BibitemOpen
  \bibfield  {author} {\bibinfo {author} {\bibfnamefont {J.}~\bibnamefont
  {Chluba}}, \bibinfo {author} {\bibfnamefont {R.}~\bibnamefont {Khatri}}, \
  and\ \bibinfo {author} {\bibfnamefont {R.~A.}\ \bibnamefont {Sunyaev}},\
  }\href {\doibase 10.1111/j.1365-2966.2012.21474.x} {\bibfield  {journal}
  {\bibinfo  {journal} {Mon. Not. Roy. Astron. Soc.}\ }\textbf {\bibinfo
  {volume} {425}},\ \bibinfo {pages} {1129} (\bibinfo {year} {2012})},\ \Eprint
  {http://arxiv.org/abs/1202.0057} {arXiv:1202.0057 [astro-ph.CO]} \BibitemShut
  {NoStop}%
\bibitem [{\citenamefont {Khatri}\ \emph {et~al.}(2012)\citenamefont {Khatri},
  \citenamefont {Sunyaev},\ and\ \citenamefont {Chluba}}]{Khatri:2012rt}%
  \BibitemOpen
  \bibfield  {author} {\bibinfo {author} {\bibfnamefont {R.}~\bibnamefont
  {Khatri}}, \bibinfo {author} {\bibfnamefont {R.~A.}\ \bibnamefont {Sunyaev}},
  \ and\ \bibinfo {author} {\bibfnamefont {J.}~\bibnamefont {Chluba}},\ }\href
  {\doibase 10.1051/0004-6361/201219590} {\bibfield  {journal} {\bibinfo
  {journal} {Astron. Astrophys.}\ }\textbf {\bibinfo {volume} {543}},\ \bibinfo
  {pages} {A136} (\bibinfo {year} {2012})},\ \Eprint
  {http://arxiv.org/abs/1205.2871} {arXiv:1205.2871 [astro-ph.CO]} \BibitemShut
  {NoStop}%
\bibitem [{\citenamefont {Khatri}\ and\ \citenamefont
  {Sunyaev}(2013)}]{Khatri:2013dha}%
  \BibitemOpen
  \bibfield  {author} {\bibinfo {author} {\bibfnamefont {R.}~\bibnamefont
  {Khatri}}\ and\ \bibinfo {author} {\bibfnamefont {R.~A.}\ \bibnamefont
  {Sunyaev}},\ }\href {\doibase 10.1088/1475-7516/2013/06/026} {\bibfield
  {journal} {\bibinfo  {journal} {JCAP}\ }\textbf {\bibinfo {volume} {1306}},\
  \bibinfo {pages} {026} (\bibinfo {year} {2013})},\ \Eprint
  {http://arxiv.org/abs/1303.7212} {arXiv:1303.7212 [astro-ph.CO]} \BibitemShut
  {NoStop}%
\bibitem [{\citenamefont {Kosowsky}\ and\ \citenamefont
  {Turner}(1995)}]{Kosowsky:1995aa}%
  \BibitemOpen
  \bibfield  {author} {\bibinfo {author} {\bibfnamefont {A.}~\bibnamefont
  {Kosowsky}}\ and\ \bibinfo {author} {\bibfnamefont {M.~S.}\ \bibnamefont
  {Turner}},\ }\href {\doibase 10.1103/PhysRevD.52.R1739} {\bibfield  {journal}
  {\bibinfo  {journal} {Phys. Rev.}\ }\textbf {\bibinfo {volume} {D52}},\
  \bibinfo {pages} {R1739} (\bibinfo {year} {1995})},\ \Eprint
  {http://arxiv.org/abs/astro-ph/9504071} {arXiv:astro-ph/9504071 [astro-ph]}
  \BibitemShut {NoStop}%
\bibitem [{\citenamefont {Mifsud}\ and\ \citenamefont {van~de
  Bruck}(2019)}]{Mifsud:PBH_PRD}%
  \BibitemOpen
  \bibfield  {author} {\bibinfo {author} {\bibfnamefont {J.}~\bibnamefont
  {Mifsud}}\ and\ \bibinfo {author} {\bibfnamefont {C.}~\bibnamefont {van~de
  Bruck}},\ }\href@noop {} {} (\bibinfo {year} {2019}),\ \bibinfo {note} {(To
  appear)}\BibitemShut {NoStop}%
\bibitem [{\citenamefont {Peebles}\ and\ \citenamefont
  {Yu}(1970)}]{Peebles:1970ag}%
  \BibitemOpen
  \bibfield  {author} {\bibinfo {author} {\bibfnamefont {P.~J.~E.}\
  \bibnamefont {Peebles}}\ and\ \bibinfo {author} {\bibfnamefont {J.~T.}\
  \bibnamefont {Yu}},\ }\href {\doibase 10.1086/150713} {\bibfield  {journal}
  {\bibinfo  {journal} {Astrophys. J.}\ }\textbf {\bibinfo {volume} {162}},\
  \bibinfo {pages} {815} (\bibinfo {year} {1970})}\BibitemShut {NoStop}%
\bibitem [{\citenamefont {Kaiser}(1983)}]{Kaiser:1983}%
  \BibitemOpen
  \bibfield  {author} {\bibinfo {author} {\bibfnamefont {N.}~\bibnamefont
  {Kaiser}},\ }\href {\doibase 10.1093/mnras/202.4.1169} {\bibfield  {journal}
  {\bibinfo  {journal} {Mon. Not. Roy. Astron. Soc.}\ }\textbf {\bibinfo
  {volume} {202}},\ \bibinfo {pages} {1169} (\bibinfo {year}
  {1983})}\BibitemShut {NoStop}%
\bibitem [{\citenamefont {Zaldarriaga}\ and\ \citenamefont
  {Harari}(1995)}]{Zaldarriaga:1995gi}%
  \BibitemOpen
  \bibfield  {author} {\bibinfo {author} {\bibfnamefont {M.}~\bibnamefont
  {Zaldarriaga}}\ and\ \bibinfo {author} {\bibfnamefont {D.~D.}\ \bibnamefont
  {Harari}},\ }\href {\doibase 10.1103/PhysRevD.52.3276} {\bibfield  {journal}
  {\bibinfo  {journal} {Phys. Rev.}\ }\textbf {\bibinfo {volume} {D52}},\
  \bibinfo {pages} {3276} (\bibinfo {year} {1995})},\ \Eprint
  {http://arxiv.org/abs/astro-ph/9504085} {arXiv:astro-ph/9504085 [astro-ph]}
  \BibitemShut {NoStop}%
\bibitem [{\citenamefont {Peebles}(1968)}]{1968ApJ...153....1P}%
  \BibitemOpen
  \bibfield  {author} {\bibinfo {author} {\bibfnamefont {P.~J.~E.}\
  \bibnamefont {Peebles}},\ }\href@noop {} {\bibfield  {journal} {\bibinfo
  {journal} {Astrophys. J.}\ }\textbf {\bibinfo {volume} {153}},\ \bibinfo
  {pages} {1} (\bibinfo {year} {1968})}\BibitemShut {NoStop}%
\bibitem [{\citenamefont {Zel'dovich}\ \emph {et~al.}(1969)\citenamefont
  {Zel'dovich}, \citenamefont {Kurt},\ and\ \citenamefont
  {Sunyaev}}]{1969JETP...28..146Z}%
  \BibitemOpen
  \bibfield  {author} {\bibinfo {author} {\bibfnamefont {Y.~B.}\ \bibnamefont
  {Zel'dovich}}, \bibinfo {author} {\bibfnamefont {V.~G.}\ \bibnamefont
  {Kurt}}, \ and\ \bibinfo {author} {\bibfnamefont {R.~A.}\ \bibnamefont
  {Sunyaev}},\ }\href@noop {} {\bibfield  {journal} {\bibinfo  {journal}
  {Soviet Journal of Experimental and Theoretical Physics}\ }\textbf {\bibinfo
  {volume} {28}},\ \bibinfo {pages} {146} (\bibinfo {year} {1969})}\BibitemShut
  {NoStop}%
\bibitem [{\citenamefont {Chluba}(2016)}]{Chluba:2016bvg}%
  \BibitemOpen
  \bibfield  {author} {\bibinfo {author} {\bibfnamefont {J.}~\bibnamefont
  {Chluba}},\ }\href {\doibase 10.1093/mnras/stw945} {\bibfield  {journal}
  {\bibinfo  {journal} {Mon. Not. Roy. Astron. Soc.}\ }\textbf {\bibinfo
  {volume} {460}},\ \bibinfo {pages} {227} (\bibinfo {year} {2016})},\ \Eprint
  {http://arxiv.org/abs/1603.02496} {arXiv:1603.02496 [astro-ph.CO]}
  \BibitemShut {NoStop}%
\bibitem [{\citenamefont {Chluba}\ \emph {et~al.}(2015)\citenamefont {Chluba},
  \citenamefont {Dai}, \citenamefont {Grin}, \citenamefont {Amin},\ and\
  \citenamefont {Kamionkowski}}]{Chluba:2014qia}%
  \BibitemOpen
  \bibfield  {author} {\bibinfo {author} {\bibfnamefont {J.}~\bibnamefont
  {Chluba}}, \bibinfo {author} {\bibfnamefont {L.}~\bibnamefont {Dai}},
  \bibinfo {author} {\bibfnamefont {D.}~\bibnamefont {Grin}}, \bibinfo {author}
  {\bibfnamefont {M.}~\bibnamefont {Amin}}, \ and\ \bibinfo {author}
  {\bibfnamefont {M.}~\bibnamefont {Kamionkowski}},\ }\href {\doibase
  10.1093/mnras/stu2277} {\bibfield  {journal} {\bibinfo  {journal} {Mon. Not.
  Roy. Astron. Soc.}\ }\textbf {\bibinfo {volume} {446}},\ \bibinfo {pages}
  {2871} (\bibinfo {year} {2015})},\ \Eprint {http://arxiv.org/abs/1407.3653}
  {arXiv:1407.3653 [astro-ph.CO]} \BibitemShut {NoStop}%
\bibitem [{\citenamefont {Chluba}\ and\ \citenamefont
  {Jeong}(2014)}]{Chluba:2013pya}%
  \BibitemOpen
  \bibfield  {author} {\bibinfo {author} {\bibfnamefont {J.}~\bibnamefont
  {Chluba}}\ and\ \bibinfo {author} {\bibfnamefont {D.}~\bibnamefont {Jeong}},\
  }\href {\doibase 10.1093/mnras/stt2327} {\bibfield  {journal} {\bibinfo
  {journal} {Mon. Not. Roy. Astron. Soc.}\ }\textbf {\bibinfo {volume} {438}},\
  \bibinfo {pages} {2065} (\bibinfo {year} {2014})},\ \Eprint
  {http://arxiv.org/abs/1306.5751} {arXiv:1306.5751 [astro-ph.CO]} \BibitemShut
  {NoStop}%
\bibitem [{\citenamefont {Khatri}\ and\ \citenamefont
  {Sunyaev}(2012{\natexlab{b}})}]{Khatri:2012tw}%
  \BibitemOpen
  \bibfield  {author} {\bibinfo {author} {\bibfnamefont {R.}~\bibnamefont
  {Khatri}}\ and\ \bibinfo {author} {\bibfnamefont {R.~A.}\ \bibnamefont
  {Sunyaev}},\ }\href {\doibase 10.1088/1475-7516/2012/09/016} {\bibfield
  {journal} {\bibinfo  {journal} {JCAP}\ }\textbf {\bibinfo {volume} {1209}},\
  \bibinfo {pages} {016} (\bibinfo {year} {2012}{\natexlab{b}})},\ \Eprint
  {http://arxiv.org/abs/1207.6654} {arXiv:1207.6654 [astro-ph.CO]} \BibitemShut
  {NoStop}%
\bibitem [{\citenamefont {Pajer}\ and\ \citenamefont
  {Zaldarriaga}(2012)}]{Pajer:2012vz}%
  \BibitemOpen
  \bibfield  {author} {\bibinfo {author} {\bibfnamefont {E.}~\bibnamefont
  {Pajer}}\ and\ \bibinfo {author} {\bibfnamefont {M.}~\bibnamefont
  {Zaldarriaga}},\ }\href {\doibase 10.1103/PhysRevLett.109.021302} {\bibfield
  {journal} {\bibinfo  {journal} {Phys. Rev. Lett.}\ }\textbf {\bibinfo
  {volume} {109}},\ \bibinfo {pages} {021302} (\bibinfo {year} {2012})},\
  \Eprint {http://arxiv.org/abs/1201.5375} {arXiv:1201.5375 [astro-ph.CO]}
  \BibitemShut {NoStop}%
\bibitem [{\citenamefont {Emami}\ \emph {et~al.}(2015)\citenamefont {Emami},
  \citenamefont {Dimastrogiovanni}, \citenamefont {Chluba},\ and\ \citenamefont
  {Kamionkowski}}]{Emami:2015xqa}%
  \BibitemOpen
  \bibfield  {author} {\bibinfo {author} {\bibfnamefont {R.}~\bibnamefont
  {Emami}}, \bibinfo {author} {\bibfnamefont {E.}~\bibnamefont
  {Dimastrogiovanni}}, \bibinfo {author} {\bibfnamefont {J.}~\bibnamefont
  {Chluba}}, \ and\ \bibinfo {author} {\bibfnamefont {M.}~\bibnamefont
  {Kamionkowski}},\ }\href {\doibase 10.1103/PhysRevD.91.123531} {\bibfield
  {journal} {\bibinfo  {journal} {Phys. Rev.}\ }\textbf {\bibinfo {volume}
  {D91}},\ \bibinfo {pages} {123531} (\bibinfo {year} {2015})},\ \Eprint
  {http://arxiv.org/abs/1504.00675} {arXiv:1504.00675 [astro-ph.CO]}
  \BibitemShut {NoStop}%
\bibitem [{\citenamefont {Chapline}(1975)}]{1975Natur.253..251C}%
  \BibitemOpen
  \bibfield  {author} {\bibinfo {author} {\bibfnamefont {G.~F.}\ \bibnamefont
  {Chapline}},\ }\href@noop {} {\bibfield  {journal} {\bibinfo  {journal}
  {Nature}\ }\textbf {\bibinfo {volume} {253}},\ \bibinfo {pages} {{251}}
  (\bibinfo {year} {1975})}\BibitemShut {NoStop}%
\bibitem [{\citenamefont {Bird}\ \emph {et~al.}(2016)\citenamefont {Bird},
  \citenamefont {Cholis}, \citenamefont {Mu{\~n}oz}, \citenamefont
  {Ali-Ha{\"i}moud}, \citenamefont {Kamionkowski}, \citenamefont {Kovetz},
  \citenamefont {Raccanelli},\ and\ \citenamefont {Riess}}]{Bird:2016dcv}%
  \BibitemOpen
  \bibfield  {author} {\bibinfo {author} {\bibfnamefont {S.}~\bibnamefont
  {Bird}}, \bibinfo {author} {\bibfnamefont {I.}~\bibnamefont {Cholis}},
  \bibinfo {author} {\bibfnamefont {J.~B.}\ \bibnamefont {Mu{\~n}oz}}, \bibinfo
  {author} {\bibfnamefont {Y.}~\bibnamefont {Ali-Ha{\"i}moud}}, \bibinfo
  {author} {\bibfnamefont {M.}~\bibnamefont {Kamionkowski}}, \bibinfo {author}
  {\bibfnamefont {E.~D.}\ \bibnamefont {Kovetz}}, \bibinfo {author}
  {\bibfnamefont {A.}~\bibnamefont {Raccanelli}}, \ and\ \bibinfo {author}
  {\bibfnamefont {A.~G.}\ \bibnamefont {Riess}},\ }\href {\doibase
  10.1103/PhysRevLett.116.201301} {\bibfield  {journal} {\bibinfo  {journal}
  {Phys. Rev. Lett.}\ }\textbf {\bibinfo {volume} {116}},\ \bibinfo {pages}
  {201301} (\bibinfo {year} {2016})},\ \Eprint
  {http://arxiv.org/abs/1603.00464} {arXiv:1603.00464 [astro-ph.CO]}
  \BibitemShut {NoStop}%
\bibitem [{\citenamefont {Abbott}\ \emph
  {et~al.}(2016{\natexlab{a}})\citenamefont {Abbott} \emph
  {et~al.}}]{Abbott:2016blz}%
  \BibitemOpen
  \bibfield  {author} {\bibinfo {author} {\bibfnamefont {B.~P.}\ \bibnamefont
  {Abbott}} \emph {et~al.} (\bibinfo {collaboration} {Virgo, LIGO
  Scientific}),\ }\href {\doibase 10.1103/PhysRevLett.116.061102} {\bibfield
  {journal} {\bibinfo  {journal} {Phys. Rev. Lett.}\ }\textbf {\bibinfo
  {volume} {116}},\ \bibinfo {pages} {061102} (\bibinfo {year}
  {2016}{\natexlab{a}})},\ \Eprint {http://arxiv.org/abs/1602.03837}
  {arXiv:1602.03837 [gr-qc]} \BibitemShut {NoStop}%
\bibitem [{\citenamefont {Belczynski}\ \emph {et~al.}(2010)\citenamefont
  {Belczynski}, \citenamefont {Bulik}, \citenamefont {Fryer}, \citenamefont
  {Ruiter}, \citenamefont {Valsecchi}, \citenamefont {Vink},\ and\
  \citenamefont {Hurley}}]{2010ApJ...714.1217B}%
  \BibitemOpen
  \bibfield  {author} {\bibinfo {author} {\bibfnamefont {K.}~\bibnamefont
  {Belczynski}}, \bibinfo {author} {\bibfnamefont {T.}~\bibnamefont {Bulik}},
  \bibinfo {author} {\bibfnamefont {C.~L.}\ \bibnamefont {Fryer}}, \bibinfo
  {author} {\bibfnamefont {A.}~\bibnamefont {Ruiter}}, \bibinfo {author}
  {\bibfnamefont {F.}~\bibnamefont {Valsecchi}}, \bibinfo {author}
  {\bibfnamefont {J.~S.}\ \bibnamefont {Vink}}, \ and\ \bibinfo {author}
  {\bibfnamefont {J.~R.}\ \bibnamefont {Hurley}},\ }\href@noop {} {\bibfield
  {journal} {\bibinfo  {journal} {Astrophys. J.}\ }\textbf {\bibinfo {volume}
  {714}},\ \bibinfo {pages} {1217} (\bibinfo {year} {2010})},\ \Eprint
  {http://arxiv.org/abs/0904.2784} {arXiv:0904.2784 [astro-ph.SR]} \BibitemShut
  {NoStop}%
\bibitem [{\citenamefont {Prestwich}\ \emph {et~al.}(2007)\citenamefont
  {Prestwich}, \citenamefont {Kilgard}, \citenamefont {Crowther}, \citenamefont
  {Carpano}, \citenamefont {Pollock}, \citenamefont {Zezas}, \citenamefont
  {Saar}, \citenamefont {Roberts},\ and\ \citenamefont
  {Ward}}]{Prestwich:2007mj}%
  \BibitemOpen
  \bibfield  {author} {\bibinfo {author} {\bibfnamefont {A.~H.}\ \bibnamefont
  {Prestwich}}, \bibinfo {author} {\bibfnamefont {R.}~\bibnamefont {Kilgard}},
  \bibinfo {author} {\bibfnamefont {P.~A.}\ \bibnamefont {Crowther}}, \bibinfo
  {author} {\bibfnamefont {S.}~\bibnamefont {Carpano}}, \bibinfo {author}
  {\bibfnamefont {A.~M.~T.}\ \bibnamefont {Pollock}}, \bibinfo {author}
  {\bibfnamefont {A.}~\bibnamefont {Zezas}}, \bibinfo {author} {\bibfnamefont
  {S.~H.}\ \bibnamefont {Saar}}, \bibinfo {author} {\bibfnamefont {T.~P.}\
  \bibnamefont {Roberts}}, \ and\ \bibinfo {author} {\bibfnamefont {M.~J.}\
  \bibnamefont {Ward}},\ }\href {\doibase 10.1086/523755} {\bibfield  {journal}
  {\bibinfo  {journal} {Astrophys. J.}\ }\textbf {\bibinfo {volume} {669}},\
  \bibinfo {pages} {L21} (\bibinfo {year} {2007})},\ \Eprint
  {http://arxiv.org/abs/0709.2892} {arXiv:0709.2892 [astro-ph]} \BibitemShut
  {NoStop}%
\bibitem [{\citenamefont {Silverman}\ and\ \citenamefont
  {Filippenko}(2008)}]{Silverman:2008ss}%
  \BibitemOpen
  \bibfield  {author} {\bibinfo {author} {\bibfnamefont {J.~M.}\ \bibnamefont
  {Silverman}}\ and\ \bibinfo {author} {\bibfnamefont {A.~V.}\ \bibnamefont
  {Filippenko}},\ }\href {\doibase 10.1086/588096} {\bibfield  {journal}
  {\bibinfo  {journal} {Astrophys. J.}\ }\textbf {\bibinfo {volume} {678}},\
  \bibinfo {pages} {L17} (\bibinfo {year} {2008})},\ \Eprint
  {http://arxiv.org/abs/0802.2716} {arXiv:0802.2716 [astro-ph]} \BibitemShut
  {NoStop}%
\bibitem [{\citenamefont {Abbott}\ \emph
  {et~al.}(2016{\natexlab{b}})\citenamefont {Abbott} \emph
  {et~al.}}]{TheLIGOScientific:2016htt}%
  \BibitemOpen
  \bibfield  {author} {\bibinfo {author} {\bibfnamefont {B.~P.}\ \bibnamefont
  {Abbott}} \emph {et~al.} (\bibinfo {collaboration} {Virgo, LIGO
  Scientific}),\ }\href {\doibase 10.3847/2041-8205/818/2/L22} {\bibfield
  {journal} {\bibinfo  {journal} {Astrophys. J.}\ }\textbf {\bibinfo {volume}
  {818}},\ \bibinfo {pages} {L22} (\bibinfo {year} {2016}{\natexlab{b}})},\
  \Eprint {http://arxiv.org/abs/1602.03846} {arXiv:1602.03846 [astro-ph.HE]}
  \BibitemShut {NoStop}%
\bibitem [{\citenamefont {Abbott}\ \emph
  {et~al.}(2016{\natexlab{c}})\citenamefont {Abbott} \emph
  {et~al.}}]{Abbott:2016nmj}%
  \BibitemOpen
  \bibfield  {author} {\bibinfo {author} {\bibfnamefont {B.~P.}\ \bibnamefont
  {Abbott}} \emph {et~al.} (\bibinfo {collaboration} {Virgo, LIGO
  Scientific}),\ }\href {\doibase 10.1103/PhysRevLett.116.241103} {\bibfield
  {journal} {\bibinfo  {journal} {Phys. Rev. Lett.}\ }\textbf {\bibinfo
  {volume} {116}},\ \bibinfo {pages} {241103} (\bibinfo {year}
  {2016}{\natexlab{c}})},\ \Eprint {http://arxiv.org/abs/1606.04855}
  {arXiv:1606.04855 [gr-qc]} \BibitemShut {NoStop}%
\bibitem [{\citenamefont {Abbott}\ \emph
  {et~al.}(2017{\natexlab{a}})\citenamefont {Abbott} \emph
  {et~al.}}]{Abbott:2017oio}%
  \BibitemOpen
  \bibfield  {author} {\bibinfo {author} {\bibfnamefont {B.~P.}\ \bibnamefont
  {Abbott}} \emph {et~al.} (\bibinfo {collaboration} {Virgo, LIGO
  Scientific}),\ }\href {\doibase 10.1103/PhysRevLett.119.141101} {\bibfield
  {journal} {\bibinfo  {journal} {Phys. Rev. Lett.}\ }\textbf {\bibinfo
  {volume} {119}},\ \bibinfo {pages} {141101} (\bibinfo {year}
  {2017}{\natexlab{a}})},\ \Eprint {http://arxiv.org/abs/1709.09660}
  {arXiv:1709.09660 [gr-qc]} \BibitemShut {NoStop}%
\bibitem [{\citenamefont {Abbott}\ \emph
  {et~al.}(2017{\natexlab{b}})\citenamefont {Abbott} \emph
  {et~al.}}]{Abbott:2017gyy}%
  \BibitemOpen
  \bibfield  {author} {\bibinfo {author} {\bibfnamefont {B.~P.}\ \bibnamefont
  {Abbott}} \emph {et~al.} (\bibinfo {collaboration} {Virgo, LIGO
  Scientific}),\ }\href {\doibase 10.3847/2041-8213/aa9f0c} {\bibfield
  {journal} {\bibinfo  {journal} {Astrophys. J.}\ }\textbf {\bibinfo {volume}
  {851}},\ \bibinfo {pages} {L35} (\bibinfo {year} {2017}{\natexlab{b}})},\
  \Eprint {http://arxiv.org/abs/1711.05578} {arXiv:1711.05578 [astro-ph.HE]}
  \BibitemShut {NoStop}%
\bibitem [{\citenamefont {Abbott}\ \emph
  {et~al.}(2017{\natexlab{c}})\citenamefont {Abbott} \emph
  {et~al.}}]{Abbott:2017vtc}%
  \BibitemOpen
  \bibfield  {author} {\bibinfo {author} {\bibfnamefont {B.~P.}\ \bibnamefont
  {Abbott}} \emph {et~al.} (\bibinfo {collaboration} {VIRGO, LIGO
  Scientific}),\ }\href {\doibase 10.1103/PhysRevLett.118.221101} {\bibfield
  {journal} {\bibinfo  {journal} {Phys. Rev. Lett.}\ }\textbf {\bibinfo
  {volume} {118}},\ \bibinfo {pages} {221101} (\bibinfo {year}
  {2017}{\natexlab{c}})},\ \Eprint {http://arxiv.org/abs/1706.01812}
  {arXiv:1706.01812 [gr-qc]} \BibitemShut {NoStop}%
\bibitem [{\citenamefont {Abbott}\ \emph {et~al.}(2018)\citenamefont {Abbott}
  \emph {et~al.}}]{LIGOScientific:2018mvr}%
  \BibitemOpen
  \bibfield  {author} {\bibinfo {author} {\bibfnamefont {B.~P.}\ \bibnamefont
  {Abbott}} \emph {et~al.} (\bibinfo {collaboration} {LIGO Scientific,
  Virgo}),\ }\href@noop {} {\bibfield  {journal} {\bibinfo  {journal} {ArXiv
  e-prints}\ } (\bibinfo {year} {2018})},\ \Eprint
  {http://arxiv.org/abs/1811.12907} {arXiv:1811.12907 [astro-ph.HE]}
  \BibitemShut {NoStop}%
\bibitem [{\citenamefont {Carr}\ \emph {et~al.}(1994)\citenamefont {Carr},
  \citenamefont {Gilbert},\ and\ \citenamefont {Lidsey}}]{Carr:1994ar}%
  \BibitemOpen
  \bibfield  {author} {\bibinfo {author} {\bibfnamefont {B.~J.}\ \bibnamefont
  {Carr}}, \bibinfo {author} {\bibfnamefont {J.~H.}\ \bibnamefont {Gilbert}}, \
  and\ \bibinfo {author} {\bibfnamefont {J.~E.}\ \bibnamefont {Lidsey}},\
  }\href {\doibase 10.1103/PhysRevD.50.4853} {\bibfield  {journal} {\bibinfo
  {journal} {Phys. Rev.}\ }\textbf {\bibinfo {volume} {D50}},\ \bibinfo {pages}
  {4853} (\bibinfo {year} {1994})},\ \Eprint
  {http://arxiv.org/abs/astro-ph/9405027} {arXiv:astro-ph/9405027 [astro-ph]}
  \BibitemShut {NoStop}%
\bibitem [{\citenamefont {Sendouda}\ \emph {et~al.}(2006)\citenamefont
  {Sendouda}, \citenamefont {Nagataki},\ and\ \citenamefont
  {Sato}}]{Sendouda:2006nu}%
  \BibitemOpen
  \bibfield  {author} {\bibinfo {author} {\bibfnamefont {Y.}~\bibnamefont
  {Sendouda}}, \bibinfo {author} {\bibfnamefont {S.}~\bibnamefont {Nagataki}},
  \ and\ \bibinfo {author} {\bibfnamefont {K.}~\bibnamefont {Sato}},\ }\href
  {\doibase 10.1088/1475-7516/2006/06/003} {\bibfield  {journal} {\bibinfo
  {journal} {JCAP}\ }\textbf {\bibinfo {volume} {0606}},\ \bibinfo {pages}
  {003} (\bibinfo {year} {2006})},\ \Eprint
  {http://arxiv.org/abs/astro-ph/0603509} {arXiv:astro-ph/0603509 [astro-ph]}
  \BibitemShut {NoStop}%
\bibitem [{\citenamefont {Carr}(1975)}]{1975ApJ...201....1C}%
  \BibitemOpen
  \bibfield  {author} {\bibinfo {author} {\bibfnamefont {B.~J.}\ \bibnamefont
  {Carr}},\ }\href@noop {} {\bibfield  {journal} {\bibinfo  {journal}
  {Astrophys. J.}\ }\textbf {\bibinfo {volume} {201}},\ \bibinfo {pages} {1}
  (\bibinfo {year} {1975})}\BibitemShut {NoStop}%
\bibitem [{\citenamefont {Polnarev}\ and\ \citenamefont
  {Musco}(2007)}]{Polnarev:2006aa}%
  \BibitemOpen
  \bibfield  {author} {\bibinfo {author} {\bibfnamefont {A.~G.}\ \bibnamefont
  {Polnarev}}\ and\ \bibinfo {author} {\bibfnamefont {I.}~\bibnamefont
  {Musco}},\ }\href {\doibase 10.1088/0264-9381/24/6/003} {\bibfield  {journal}
  {\bibinfo  {journal} {Class. Quant. Grav.}\ }\textbf {\bibinfo {volume}
  {24}},\ \bibinfo {pages} {1405} (\bibinfo {year} {2007})},\ \Eprint
  {http://arxiv.org/abs/gr-qc/0605122} {arXiv:gr-qc/0605122 [gr-qc]}
  \BibitemShut {NoStop}%
\bibitem [{\citenamefont {Musco}\ \emph {et~al.}(2005)\citenamefont {Musco},
  \citenamefont {Miller},\ and\ \citenamefont {Rezzolla}}]{Musco:2004ak}%
  \BibitemOpen
  \bibfield  {author} {\bibinfo {author} {\bibfnamefont {I.}~\bibnamefont
  {Musco}}, \bibinfo {author} {\bibfnamefont {J.~C.}\ \bibnamefont {Miller}}, \
  and\ \bibinfo {author} {\bibfnamefont {L.}~\bibnamefont {Rezzolla}},\ }\href
  {\doibase 10.1088/0264-9381/22/7/013} {\bibfield  {journal} {\bibinfo
  {journal} {Class. Quant. Grav.}\ }\textbf {\bibinfo {volume} {22}},\ \bibinfo
  {pages} {1405} (\bibinfo {year} {2005})},\ \Eprint
  {http://arxiv.org/abs/gr-qc/0412063} {arXiv:gr-qc/0412063 [gr-qc]}
  \BibitemShut {NoStop}%
\bibitem [{\citenamefont {Harada}\ and\ \citenamefont
  {Carr}(2005)}]{Harada:2004pe}%
  \BibitemOpen
  \bibfield  {author} {\bibinfo {author} {\bibfnamefont {T.}~\bibnamefont
  {Harada}}\ and\ \bibinfo {author} {\bibfnamefont {B.~J.}\ \bibnamefont
  {Carr}},\ }\href {\doibase 10.1103/PhysRevD.71.104009} {\bibfield  {journal}
  {\bibinfo  {journal} {Phys. Rev.}\ }\textbf {\bibinfo {volume} {D71}},\
  \bibinfo {pages} {104009} (\bibinfo {year} {2005})},\ \Eprint
  {http://arxiv.org/abs/astro-ph/0412134} {arXiv:astro-ph/0412134 [astro-ph]}
  \BibitemShut {NoStop}%
\bibitem [{\citenamefont {Kopp}\ \emph {et~al.}(2011)\citenamefont {Kopp},
  \citenamefont {Hofmann},\ and\ \citenamefont {Weller}}]{Kopp:2010sh}%
  \BibitemOpen
  \bibfield  {author} {\bibinfo {author} {\bibfnamefont {M.}~\bibnamefont
  {Kopp}}, \bibinfo {author} {\bibfnamefont {S.}~\bibnamefont {Hofmann}}, \
  and\ \bibinfo {author} {\bibfnamefont {J.}~\bibnamefont {Weller}},\ }\href
  {\doibase 10.1103/PhysRevD.83.124025} {\bibfield  {journal} {\bibinfo
  {journal} {Phys. Rev.}\ }\textbf {\bibinfo {volume} {D83}},\ \bibinfo {pages}
  {124025} (\bibinfo {year} {2011})},\ \Eprint {http://arxiv.org/abs/1012.4369}
  {arXiv:1012.4369 [astro-ph.CO]} \BibitemShut {NoStop}%
\bibitem [{\citenamefont {Blais}\ \emph {et~al.}(2003)\citenamefont {Blais},
  \citenamefont {Bringmann}, \citenamefont {Kiefer},\ and\ \citenamefont
  {Polarski}}]{Blais:2002gw}%
  \BibitemOpen
  \bibfield  {author} {\bibinfo {author} {\bibfnamefont {D.}~\bibnamefont
  {Blais}}, \bibinfo {author} {\bibfnamefont {T.}~\bibnamefont {Bringmann}},
  \bibinfo {author} {\bibfnamefont {C.}~\bibnamefont {Kiefer}}, \ and\ \bibinfo
  {author} {\bibfnamefont {D.}~\bibnamefont {Polarski}},\ }\href {\doibase
  10.1103/PhysRevD.67.024024} {\bibfield  {journal} {\bibinfo  {journal} {Phys.
  Rev.}\ }\textbf {\bibinfo {volume} {D67}},\ \bibinfo {pages} {024024}
  (\bibinfo {year} {2003})},\ \Eprint {http://arxiv.org/abs/astro-ph/0206262}
  {arXiv:astro-ph/0206262 [astro-ph]} \BibitemShut {NoStop}%
\bibitem [{\citenamefont {Bringmann}\ \emph {et~al.}(2002)\citenamefont
  {Bringmann}, \citenamefont {Kiefer},\ and\ \citenamefont
  {Polarski}}]{Bringmann:2001yp}%
  \BibitemOpen
  \bibfield  {author} {\bibinfo {author} {\bibfnamefont {T.}~\bibnamefont
  {Bringmann}}, \bibinfo {author} {\bibfnamefont {C.}~\bibnamefont {Kiefer}}, \
  and\ \bibinfo {author} {\bibfnamefont {D.}~\bibnamefont {Polarski}},\ }\href
  {\doibase 10.1103/PhysRevD.65.024008} {\bibfield  {journal} {\bibinfo
  {journal} {Phys. Rev.}\ }\textbf {\bibinfo {volume} {D65}},\ \bibinfo {pages}
  {024008} (\bibinfo {year} {2002})},\ \Eprint
  {http://arxiv.org/abs/astro-ph/0109404} {arXiv:astro-ph/0109404 [astro-ph]}
  \BibitemShut {NoStop}%
\bibitem [{\citenamefont {Josan}\ \emph {et~al.}(2009)\citenamefont {Josan},
  \citenamefont {Green},\ and\ \citenamefont {Malik}}]{Josan:2009qn}%
  \BibitemOpen
  \bibfield  {author} {\bibinfo {author} {\bibfnamefont {A.~S.}\ \bibnamefont
  {Josan}}, \bibinfo {author} {\bibfnamefont {A.~M.}\ \bibnamefont {Green}}, \
  and\ \bibinfo {author} {\bibfnamefont {K.~A.}\ \bibnamefont {Malik}},\ }\href
  {\doibase 10.1103/PhysRevD.79.103520} {\bibfield  {journal} {\bibinfo
  {journal} {Phys. Rev.}\ }\textbf {\bibinfo {volume} {D79}},\ \bibinfo {pages}
  {103520} (\bibinfo {year} {2009})},\ \Eprint {http://arxiv.org/abs/0903.3184}
  {arXiv:0903.3184 [astro-ph.CO]} \BibitemShut {NoStop}%
\bibitem [{\citenamefont {Drees}\ and\ \citenamefont
  {Erfani}(2012)}]{Drees:2011yz}%
  \BibitemOpen
  \bibfield  {author} {\bibinfo {author} {\bibfnamefont {M.}~\bibnamefont
  {Drees}}\ and\ \bibinfo {author} {\bibfnamefont {E.}~\bibnamefont {Erfani}},\
  }\href {\doibase 10.1088/1475-7516/2012/01/035} {\bibfield  {journal}
  {\bibinfo  {journal} {JCAP}\ }\textbf {\bibinfo {volume} {1201}},\ \bibinfo
  {pages} {035} (\bibinfo {year} {2012})},\ \Eprint
  {http://arxiv.org/abs/1110.6052} {arXiv:1110.6052 [astro-ph.CO]} \BibitemShut
  {NoStop}%
\bibitem [{\citenamefont {Drees}\ and\ \citenamefont
  {Erfani}(2011)}]{Drees:2011hb}%
  \BibitemOpen
  \bibfield  {author} {\bibinfo {author} {\bibfnamefont {M.}~\bibnamefont
  {Drees}}\ and\ \bibinfo {author} {\bibfnamefont {E.}~\bibnamefont {Erfani}},\
  }\href {\doibase 10.1088/1475-7516/2011/04/005} {\bibfield  {journal}
  {\bibinfo  {journal} {JCAP}\ }\textbf {\bibinfo {volume} {1104}},\ \bibinfo
  {pages} {005} (\bibinfo {year} {2011})},\ \Eprint
  {http://arxiv.org/abs/1102.2340} {arXiv:1102.2340 [hep-ph]} \BibitemShut
  {NoStop}%
\bibitem [{\citenamefont {Press}\ and\ \citenamefont
  {Schechter}(1974)}]{Press:1973iz}%
  \BibitemOpen
  \bibfield  {author} {\bibinfo {author} {\bibfnamefont {W.~H.}\ \bibnamefont
  {Press}}\ and\ \bibinfo {author} {\bibfnamefont {P.}~\bibnamefont
  {Schechter}},\ }\href {\doibase 10.1086/152650} {\bibfield  {journal}
  {\bibinfo  {journal} {Astrophys. J.}\ }\textbf {\bibinfo {volume} {187}},\
  \bibinfo {pages} {425} (\bibinfo {year} {1974})}\BibitemShut {NoStop}%
\bibitem [{\citenamefont {Green}\ \emph {et~al.}(2004)\citenamefont {Green},
  \citenamefont {Liddle}, \citenamefont {Malik},\ and\ \citenamefont
  {Sasaki}}]{Green:2004wb}%
  \BibitemOpen
  \bibfield  {author} {\bibinfo {author} {\bibfnamefont {A.~M.}\ \bibnamefont
  {Green}}, \bibinfo {author} {\bibfnamefont {A.~R.}\ \bibnamefont {Liddle}},
  \bibinfo {author} {\bibfnamefont {K.~A.}\ \bibnamefont {Malik}}, \ and\
  \bibinfo {author} {\bibfnamefont {M.}~\bibnamefont {Sasaki}},\ }\href
  {\doibase 10.1103/PhysRevD.70.041502} {\bibfield  {journal} {\bibinfo
  {journal} {Phys. Rev.}\ }\textbf {\bibinfo {volume} {D70}},\ \bibinfo {pages}
  {041502} (\bibinfo {year} {2004})},\ \Eprint
  {http://arxiv.org/abs/astro-ph/0403181} {arXiv:astro-ph/0403181 [astro-ph]}
  \BibitemShut {NoStop}%
\bibitem [{\citenamefont {Carr}\ \emph {et~al.}(2010)\citenamefont {Carr},
  \citenamefont {Kohri}, \citenamefont {Sendouda},\ and\ \citenamefont
  {Yokoyama}}]{Carr:2009jm}%
  \BibitemOpen
  \bibfield  {author} {\bibinfo {author} {\bibfnamefont {B.~J.}\ \bibnamefont
  {Carr}}, \bibinfo {author} {\bibfnamefont {K.}~\bibnamefont {Kohri}},
  \bibinfo {author} {\bibfnamefont {Y.}~\bibnamefont {Sendouda}}, \ and\
  \bibinfo {author} {\bibfnamefont {J.}~\bibnamefont {Yokoyama}},\ }\href
  {\doibase 10.1103/PhysRevD.81.104019} {\bibfield  {journal} {\bibinfo
  {journal} {Phys. Rev.}\ }\textbf {\bibinfo {volume} {D81}},\ \bibinfo {pages}
  {104019} (\bibinfo {year} {2010})},\ \Eprint {http://arxiv.org/abs/0912.5297}
  {arXiv:0912.5297 [astro-ph.CO]} \BibitemShut {NoStop}%
\bibitem [{\citenamefont {Barnacka}\ \emph {et~al.}(2012)\citenamefont
  {Barnacka}, \citenamefont {Glicenstein},\ and\ \citenamefont
  {Moderski}}]{Barnacka:2012bm}%
  \BibitemOpen
  \bibfield  {author} {\bibinfo {author} {\bibfnamefont {A.}~\bibnamefont
  {Barnacka}}, \bibinfo {author} {\bibfnamefont {J.~F.}\ \bibnamefont
  {Glicenstein}}, \ and\ \bibinfo {author} {\bibfnamefont {R.}~\bibnamefont
  {Moderski}},\ }\href {\doibase 10.1103/PhysRevD.86.043001} {\bibfield
  {journal} {\bibinfo  {journal} {Phys. Rev.}\ }\textbf {\bibinfo {volume}
  {D86}},\ \bibinfo {pages} {043001} (\bibinfo {year} {2012})},\ \Eprint
  {http://arxiv.org/abs/1204.2056} {arXiv:1204.2056 [astro-ph.CO]} \BibitemShut
  {NoStop}%
\bibitem [{\citenamefont {Graham}\ \emph {et~al.}(2015)\citenamefont {Graham},
  \citenamefont {Rajendran},\ and\ \citenamefont {Varela}}]{Graham:2015apa}%
  \BibitemOpen
  \bibfield  {author} {\bibinfo {author} {\bibfnamefont {P.~W.}\ \bibnamefont
  {Graham}}, \bibinfo {author} {\bibfnamefont {S.}~\bibnamefont {Rajendran}}, \
  and\ \bibinfo {author} {\bibfnamefont {J.}~\bibnamefont {Varela}},\ }\href
  {\doibase 10.1103/PhysRevD.92.063007} {\bibfield  {journal} {\bibinfo
  {journal} {Phys. Rev.}\ }\textbf {\bibinfo {volume} {D92}},\ \bibinfo {pages}
  {063007} (\bibinfo {year} {2015})},\ \Eprint
  {http://arxiv.org/abs/1505.04444} {arXiv:1505.04444 [hep-ph]} \BibitemShut
  {NoStop}%
\bibitem [{\citenamefont {Capela}\ \emph {et~al.}(2013)\citenamefont {Capela},
  \citenamefont {Pshirkov},\ and\ \citenamefont {Tinyakov}}]{Capela:2013yf}%
  \BibitemOpen
  \bibfield  {author} {\bibinfo {author} {\bibfnamefont {F.}~\bibnamefont
  {Capela}}, \bibinfo {author} {\bibfnamefont {M.}~\bibnamefont {Pshirkov}}, \
  and\ \bibinfo {author} {\bibfnamefont {P.}~\bibnamefont {Tinyakov}},\ }\href
  {\doibase 10.1103/PhysRevD.87.123524} {\bibfield  {journal} {\bibinfo
  {journal} {Phys. Rev.}\ }\textbf {\bibinfo {volume} {D87}},\ \bibinfo {pages}
  {123524} (\bibinfo {year} {2013})},\ \Eprint {http://arxiv.org/abs/1301.4984}
  {arXiv:1301.4984 [astro-ph.CO]} \BibitemShut {NoStop}%
\bibitem [{\citenamefont {Niikura}\ \emph {et~al.}(2017)\citenamefont
  {Niikura}, \citenamefont {Takada}, \citenamefont {Yasuda}, \citenamefont
  {Lupton}, \citenamefont {Sumi}, \citenamefont {More}, \citenamefont {More},
  \citenamefont {Oguri},\ and\ \citenamefont {Chiba}}]{Niikura:2017zjd}%
  \BibitemOpen
  \bibfield  {author} {\bibinfo {author} {\bibfnamefont {H.}~\bibnamefont
  {Niikura}}, \bibinfo {author} {\bibfnamefont {M.}~\bibnamefont {Takada}},
  \bibinfo {author} {\bibfnamefont {N.}~\bibnamefont {Yasuda}}, \bibinfo
  {author} {\bibfnamefont {R.~H.}\ \bibnamefont {Lupton}}, \bibinfo {author}
  {\bibfnamefont {T.}~\bibnamefont {Sumi}}, \bibinfo {author} {\bibfnamefont
  {S.}~\bibnamefont {More}}, \bibinfo {author} {\bibfnamefont {A.}~\bibnamefont
  {More}}, \bibinfo {author} {\bibfnamefont {M.}~\bibnamefont {Oguri}}, \ and\
  \bibinfo {author} {\bibfnamefont {M.}~\bibnamefont {Chiba}},\ }\href@noop {}
  {\bibfield  {journal} {\bibinfo  {journal} {ArXiv e-prints}\ } (\bibinfo
  {year} {2017})},\ \Eprint {http://arxiv.org/abs/1701.02151} {arXiv:1701.02151
  [astro-ph.CO]} \BibitemShut {NoStop}%
\bibitem [{\citenamefont {Griest}\ \emph {et~al.}(2014)\citenamefont {Griest},
  \citenamefont {Cieplak},\ and\ \citenamefont {Lehner}}]{Griest:2013aaa}%
  \BibitemOpen
  \bibfield  {author} {\bibinfo {author} {\bibfnamefont {K.}~\bibnamefont
  {Griest}}, \bibinfo {author} {\bibfnamefont {A.~M.}\ \bibnamefont {Cieplak}},
  \ and\ \bibinfo {author} {\bibfnamefont {M.~J.}\ \bibnamefont {Lehner}},\
  }\href {\doibase 10.1088/0004-637X/786/2/158} {\bibfield  {journal} {\bibinfo
   {journal} {Astrophys. J.}\ }\textbf {\bibinfo {volume} {786}},\ \bibinfo
  {pages} {158} (\bibinfo {year} {2014})},\ \Eprint
  {http://arxiv.org/abs/1307.5798} {arXiv:1307.5798 [astro-ph.CO]} \BibitemShut
  {NoStop}%
\bibitem [{\citenamefont {Tisserand}\ \emph {et~al.}(2007)\citenamefont
  {Tisserand} \emph {et~al.}}]{Tisserand:2006zx}%
  \BibitemOpen
  \bibfield  {author} {\bibinfo {author} {\bibfnamefont {P.}~\bibnamefont
  {Tisserand}} \emph {et~al.} (\bibinfo {collaboration} {EROS--2}),\ }\href
  {\doibase 10.1051/0004-6361:20066017} {\bibfield  {journal} {\bibinfo
  {journal} {Astron. Astrophys.}\ }\textbf {\bibinfo {volume} {469}},\ \bibinfo
  {pages} {387} (\bibinfo {year} {2007})},\ \Eprint
  {http://arxiv.org/abs/astro-ph/0607207} {arXiv:astro-ph/0607207 [astro-ph]}
  \BibitemShut {NoStop}%
\bibitem [{\citenamefont {Oguri}\ \emph {et~al.}(2018)\citenamefont {Oguri},
  \citenamefont {Diego}, \citenamefont {Kaiser}, \citenamefont {Kelly},\ and\
  \citenamefont {Broadhurst}}]{Oguri:2017ock}%
  \BibitemOpen
  \bibfield  {author} {\bibinfo {author} {\bibfnamefont {M.}~\bibnamefont
  {Oguri}}, \bibinfo {author} {\bibfnamefont {J.~M.}\ \bibnamefont {Diego}},
  \bibinfo {author} {\bibfnamefont {N.}~\bibnamefont {Kaiser}}, \bibinfo
  {author} {\bibfnamefont {P.~L.}\ \bibnamefont {Kelly}}, \ and\ \bibinfo
  {author} {\bibfnamefont {T.}~\bibnamefont {Broadhurst}},\ }\href {\doibase
  10.1103/PhysRevD.97.023518} {\bibfield  {journal} {\bibinfo  {journal} {Phys.
  Rev.}\ }\textbf {\bibinfo {volume} {D97}},\ \bibinfo {pages} {023518}
  (\bibinfo {year} {2018})},\ \Eprint {http://arxiv.org/abs/1710.00148}
  {arXiv:1710.00148 [astro-ph.CO]} \BibitemShut {NoStop}%
\bibitem [{\citenamefont {Allsman}\ \emph {et~al.}(2001)\citenamefont {Allsman}
  \emph {et~al.}}]{Allsman:2000kg}%
  \BibitemOpen
  \bibfield  {author} {\bibinfo {author} {\bibfnamefont {R.~A.}\ \bibnamefont
  {Allsman}} \emph {et~al.} (\bibinfo {collaboration} {Macho}),\ }\href
  {\doibase 10.1086/319636} {\bibfield  {journal} {\bibinfo  {journal}
  {Astrophys. J.}\ }\textbf {\bibinfo {volume} {550}},\ \bibinfo {pages} {L169}
  (\bibinfo {year} {2001})},\ \Eprint {http://arxiv.org/abs/astro-ph/0011506}
  {arXiv:astro-ph/0011506 [astro-ph]} \BibitemShut {NoStop}%
\bibitem [{\citenamefont {Quinn}\ \emph {et~al.}(2009)\citenamefont {Quinn},
  \citenamefont {Wilkinson}, \citenamefont {Irwin}, \citenamefont {Marshall},
  \citenamefont {Koch},\ and\ \citenamefont {Belokurov}}]{Quinn:2009zg}%
  \BibitemOpen
  \bibfield  {author} {\bibinfo {author} {\bibfnamefont {D.~P.}\ \bibnamefont
  {Quinn}}, \bibinfo {author} {\bibfnamefont {M.~I.}\ \bibnamefont
  {Wilkinson}}, \bibinfo {author} {\bibfnamefont {M.~J.}\ \bibnamefont
  {Irwin}}, \bibinfo {author} {\bibfnamefont {J.}~\bibnamefont {Marshall}},
  \bibinfo {author} {\bibfnamefont {A.}~\bibnamefont {Koch}}, \ and\ \bibinfo
  {author} {\bibfnamefont {V.}~\bibnamefont {Belokurov}},\ }\href {\doibase
  10.1111/j.1745-3933.2009.00652.x} {\bibfield  {journal} {\bibinfo  {journal}
  {Mon. Not. Roy. Astron. Soc.}\ }\textbf {\bibinfo {volume} {396}},\ \bibinfo
  {pages} {11} (\bibinfo {year} {2009})},\ \Eprint
  {http://arxiv.org/abs/0903.1644} {arXiv:0903.1644 [astro-ph.GA]} \BibitemShut
  {NoStop}%
\bibitem [{\citenamefont {Monroy-Rodr\'{\i}guez}\ and\ \citenamefont
  {Allen}(2014)}]{Monroy-Rodriguez:2014ula}%
  \BibitemOpen
  \bibfield  {author} {\bibinfo {author} {\bibfnamefont {M.~A.}\ \bibnamefont
  {Monroy-Rodr\'{\i}guez}}\ and\ \bibinfo {author} {\bibfnamefont
  {C.}~\bibnamefont {Allen}},\ }\href {\doibase 10.1088/0004-637X/790/2/159}
  {\bibfield  {journal} {\bibinfo  {journal} {Astrophys. J.}\ }\textbf
  {\bibinfo {volume} {790}},\ \bibinfo {pages} {159} (\bibinfo {year}
  {2014})},\ \Eprint {http://arxiv.org/abs/1406.5169} {arXiv:1406.5169
  [astro-ph.GA]} \BibitemShut {NoStop}%
\bibitem [{\citenamefont {Koushiappas}\ and\ \citenamefont
  {Loeb}(2017)}]{Koushiappas:2017chw}%
  \BibitemOpen
  \bibfield  {author} {\bibinfo {author} {\bibfnamefont {S.~M.}\ \bibnamefont
  {Koushiappas}}\ and\ \bibinfo {author} {\bibfnamefont {A.}~\bibnamefont
  {Loeb}},\ }\href {\doibase 10.1103/PhysRevLett.119.041102} {\bibfield
  {journal} {\bibinfo  {journal} {Phys. Rev. Lett.}\ }\textbf {\bibinfo
  {volume} {119}},\ \bibinfo {pages} {041102} (\bibinfo {year} {2017})},\
  \Eprint {http://arxiv.org/abs/1704.01668} {arXiv:1704.01668 [astro-ph.GA]}
  \BibitemShut {NoStop}%
\bibitem [{\citenamefont {Ali-Ha{\"i}moud}\ and\ \citenamefont
  {Kamionkowski}(2017)}]{Ali-Haimoud:2016mbv}%
  \BibitemOpen
  \bibfield  {author} {\bibinfo {author} {\bibfnamefont {Y.}~\bibnamefont
  {Ali-Ha{\"i}moud}}\ and\ \bibinfo {author} {\bibfnamefont {M.}~\bibnamefont
  {Kamionkowski}},\ }\href {\doibase 10.1103/PhysRevD.95.043534} {\bibfield
  {journal} {\bibinfo  {journal} {Phys. Rev.}\ }\textbf {\bibinfo {volume}
  {D95}},\ \bibinfo {pages} {043534} (\bibinfo {year} {2017})},\ \Eprint
  {http://arxiv.org/abs/1612.05644} {arXiv:1612.05644 [astro-ph.CO]}
  \BibitemShut {NoStop}%
\bibitem [{\citenamefont {Brandt}(2016)}]{Brandt:2016aco}%
  \BibitemOpen
  \bibfield  {author} {\bibinfo {author} {\bibfnamefont {T.~D.}\ \bibnamefont
  {Brandt}},\ }\href {\doibase 10.3847/2041-8205/824/2/L31} {\bibfield
  {journal} {\bibinfo  {journal} {Astrophys. J.}\ }\textbf {\bibinfo {volume}
  {824}},\ \bibinfo {pages} {L31} (\bibinfo {year} {2016})},\ \Eprint
  {http://arxiv.org/abs/1605.03665} {arXiv:1605.03665 [astro-ph.GA]}
  \BibitemShut {NoStop}%
\bibitem [{\citenamefont {Inoue}\ and\ \citenamefont
  {Kusenko}(2017)}]{Inoue:2017csr}%
  \BibitemOpen
  \bibfield  {author} {\bibinfo {author} {\bibfnamefont {Y.}~\bibnamefont
  {Inoue}}\ and\ \bibinfo {author} {\bibfnamefont {A.}~\bibnamefont
  {Kusenko}},\ }\href {\doibase 10.1088/1475-7516/2017/10/034} {\bibfield
  {journal} {\bibinfo  {journal} {JCAP}\ }\textbf {\bibinfo {volume} {1710}},\
  \bibinfo {pages} {034} (\bibinfo {year} {2017})},\ \Eprint
  {http://arxiv.org/abs/1705.00791} {arXiv:1705.00791 [astro-ph.CO]}
  \BibitemShut {NoStop}%
\bibitem [{\citenamefont {Wilkinson}\ \emph {et~al.}(2001)\citenamefont
  {Wilkinson}, \citenamefont {Henstock}, \citenamefont {Browne}, \citenamefont
  {Polatidis}, \citenamefont {Augusto}, \citenamefont {Readhead}, \citenamefont
  {Pearson}, \citenamefont {Xu}, \citenamefont {Taylor},\ and\ \citenamefont
  {Vermeulen}}]{Wilkinson:2001vv}%
  \BibitemOpen
  \bibfield  {author} {\bibinfo {author} {\bibfnamefont {P.~N.}\ \bibnamefont
  {Wilkinson}}, \bibinfo {author} {\bibfnamefont {D.~R.}\ \bibnamefont
  {Henstock}}, \bibinfo {author} {\bibfnamefont {I.~W.~A.}\ \bibnamefont
  {Browne}}, \bibinfo {author} {\bibfnamefont {A.~G.}\ \bibnamefont
  {Polatidis}}, \bibinfo {author} {\bibfnamefont {P.}~\bibnamefont {Augusto}},
  \bibinfo {author} {\bibfnamefont {A.~C.~S.}\ \bibnamefont {Readhead}},
  \bibinfo {author} {\bibfnamefont {T.~J.}\ \bibnamefont {Pearson}}, \bibinfo
  {author} {\bibfnamefont {W.}~\bibnamefont {Xu}}, \bibinfo {author}
  {\bibfnamefont {G.~B.}\ \bibnamefont {Taylor}}, \ and\ \bibinfo {author}
  {\bibfnamefont {R.~C.}\ \bibnamefont {Vermeulen}},\ }\href {\doibase
  10.1103/PhysRevLett.86.584} {\bibfield  {journal} {\bibinfo  {journal} {Phys.
  Rev. Lett.}\ }\textbf {\bibinfo {volume} {86}},\ \bibinfo {pages} {584}
  (\bibinfo {year} {2001})},\ \Eprint {http://arxiv.org/abs/astro-ph/0101328}
  {arXiv:astro-ph/0101328 [astro-ph]} \BibitemShut {NoStop}%
\bibitem [{\citenamefont {Carr}\ \emph
  {et~al.}(2016{\natexlab{a}})\citenamefont {Carr}, \citenamefont {Kuhnel},\
  and\ \citenamefont {Sandstad}}]{Carr:2016drx}%
  \BibitemOpen
  \bibfield  {author} {\bibinfo {author} {\bibfnamefont {B.}~\bibnamefont
  {Carr}}, \bibinfo {author} {\bibfnamefont {F.}~\bibnamefont {Kuhnel}}, \ and\
  \bibinfo {author} {\bibfnamefont {M.}~\bibnamefont {Sandstad}},\ }\href
  {\doibase 10.1103/PhysRevD.94.083504} {\bibfield  {journal} {\bibinfo
  {journal} {Phys. Rev.}\ }\textbf {\bibinfo {volume} {D94}},\ \bibinfo {pages}
  {083504} (\bibinfo {year} {2016}{\natexlab{a}})},\ \Eprint
  {http://arxiv.org/abs/1607.06077} {arXiv:1607.06077 [astro-ph.CO]}
  \BibitemShut {NoStop}%
\bibitem [{\citenamefont {Khlopov}(2015)}]{Khlopov:2015oda}%
  \BibitemOpen
  \bibfield  {author} {\bibinfo {author} {\bibfnamefont {M.}~\bibnamefont
  {Khlopov}},\ }\href {\doibase 10.3390/sym7020815} {\bibfield  {journal}
  {\bibinfo  {journal} {Symmetry}\ }\textbf {\bibinfo {volume} {7}},\ \bibinfo
  {pages} {815} (\bibinfo {year} {2015})},\ \Eprint
  {http://arxiv.org/abs/1505.08077} {arXiv:1505.08077 [astro-ph.CO]}
  \BibitemShut {NoStop}%
\bibitem [{\citenamefont {Lehoucq}\ \emph {et~al.}(2009)\citenamefont
  {Lehoucq}, \citenamefont {Casse}, \citenamefont {Casandjian},\ and\
  \citenamefont {Grenier}}]{Lehoucq:2009ge}%
  \BibitemOpen
  \bibfield  {author} {\bibinfo {author} {\bibfnamefont {R.}~\bibnamefont
  {Lehoucq}}, \bibinfo {author} {\bibfnamefont {M.}~\bibnamefont {Casse}},
  \bibinfo {author} {\bibfnamefont {J.~M.}\ \bibnamefont {Casandjian}}, \ and\
  \bibinfo {author} {\bibfnamefont {I.}~\bibnamefont {Grenier}},\ }\href
  {\doibase 10.1051/0004-6361/200911961} {\bibfield  {journal} {\bibinfo
  {journal} {Astron. Astrophys.}\ }\textbf {\bibinfo {volume} {502}},\ \bibinfo
  {pages} {37} (\bibinfo {year} {2009})},\ \Eprint
  {http://arxiv.org/abs/0906.1648} {arXiv:0906.1648 [astro-ph.HE]} \BibitemShut
  {NoStop}%
\bibitem [{\citenamefont {Carr}\ \emph
  {et~al.}(2016{\natexlab{b}})\citenamefont {Carr}, \citenamefont {Kohri},
  \citenamefont {Sendouda},\ and\ \citenamefont {Yokoyama}}]{Carr:2016hva}%
  \BibitemOpen
  \bibfield  {author} {\bibinfo {author} {\bibfnamefont {B.~J.}\ \bibnamefont
  {Carr}}, \bibinfo {author} {\bibfnamefont {K.}~\bibnamefont {Kohri}},
  \bibinfo {author} {\bibfnamefont {Y.}~\bibnamefont {Sendouda}}, \ and\
  \bibinfo {author} {\bibfnamefont {J.}~\bibnamefont {Yokoyama}},\ }\href
  {\doibase 10.1103/PhysRevD.94.044029} {\bibfield  {journal} {\bibinfo
  {journal} {Phys. Rev.}\ }\textbf {\bibinfo {volume} {D94}},\ \bibinfo {pages}
  {044029} (\bibinfo {year} {2016}{\natexlab{b}})},\ \Eprint
  {http://arxiv.org/abs/1604.05349} {arXiv:1604.05349 [astro-ph.CO]}
  \BibitemShut {NoStop}%
\bibitem [{\citenamefont {Mack}\ and\ \citenamefont
  {Wesley}(2008)}]{Mack:2008nv}%
  \BibitemOpen
  \bibfield  {author} {\bibinfo {author} {\bibfnamefont {K.~J.}\ \bibnamefont
  {Mack}}\ and\ \bibinfo {author} {\bibfnamefont {D.~H.}\ \bibnamefont
  {Wesley}},\ }\href@noop {} {\bibfield  {journal} {\bibinfo  {journal} {ArXiv
  e-prints}\ } (\bibinfo {year} {2008})},\ \Eprint
  {http://arxiv.org/abs/0805.1531} {arXiv:0805.1531 [astro-ph]} \BibitemShut
  {NoStop}%
\bibitem [{\citenamefont {Blas}\ \emph {et~al.}(2011)\citenamefont {Blas},
  \citenamefont {Lesgourgues},\ and\ \citenamefont {Tram}}]{Blas:2011rf}%
  \BibitemOpen
  \bibfield  {author} {\bibinfo {author} {\bibfnamefont {D.}~\bibnamefont
  {Blas}}, \bibinfo {author} {\bibfnamefont {J.}~\bibnamefont {Lesgourgues}}, \
  and\ \bibinfo {author} {\bibfnamefont {T.}~\bibnamefont {Tram}},\ }\href
  {\doibase 10.1088/1475-7516/2011/07/034} {\bibfield  {journal} {\bibinfo
  {journal} {JCAP}\ }\textbf {\bibinfo {volume} {1107}},\ \bibinfo {pages}
  {034} (\bibinfo {year} {2011})},\ \Eprint {http://arxiv.org/abs/1104.2933}
  {arXiv:1104.2933 [astro-ph.CO]} \BibitemShut {NoStop}%
\bibitem [{\citenamefont {Audren}\ \emph {et~al.}(2013)\citenamefont {Audren},
  \citenamefont {Lesgourgues}, \citenamefont {Benabed},\ and\ \citenamefont
  {Prunet}}]{Audren:2012wb}%
  \BibitemOpen
  \bibfield  {author} {\bibinfo {author} {\bibfnamefont {B.}~\bibnamefont
  {Audren}}, \bibinfo {author} {\bibfnamefont {J.}~\bibnamefont {Lesgourgues}},
  \bibinfo {author} {\bibfnamefont {K.}~\bibnamefont {Benabed}}, \ and\
  \bibinfo {author} {\bibfnamefont {S.}~\bibnamefont {Prunet}},\ }\href
  {\doibase 10.1088/1475-7516/2013/02/001} {\bibfield  {journal} {\bibinfo
  {journal} {JCAP}\ }\textbf {\bibinfo {volume} {1302}},\ \bibinfo {pages}
  {001} (\bibinfo {year} {2013})},\ \Eprint {http://arxiv.org/abs/1210.7183}
  {arXiv:1210.7183 [astro-ph.CO]} \BibitemShut {NoStop}%
\bibitem [{\citenamefont {Lewis}\ and\ \citenamefont
  {Bridle}(2002)}]{Lewis:2002ah}%
  \BibitemOpen
  \bibfield  {author} {\bibinfo {author} {\bibfnamefont {A.}~\bibnamefont
  {Lewis}}\ and\ \bibinfo {author} {\bibfnamefont {S.}~\bibnamefont {Bridle}},\
  }\href {\doibase 10.1103/PhysRevD.66.103511} {\bibfield  {journal} {\bibinfo
  {journal} {Phys. Rev.}\ }\textbf {\bibinfo {volume} {D66}},\ \bibinfo {pages}
  {103511} (\bibinfo {year} {2002})},\ \Eprint
  {http://arxiv.org/abs/astro-ph/0205436} {arXiv:astro-ph/0205436 [astro-ph]}
  \BibitemShut {NoStop}%
\bibitem [{\citenamefont {Ali-Ha{\"i}moud}\ and\ \citenamefont
  {Hirata}(2011)}]{PhysRevD.83.043513}%
  \BibitemOpen
  \bibfield  {author} {\bibinfo {author} {\bibfnamefont {Y.}~\bibnamefont
  {Ali-Ha{\"i}moud}}\ and\ \bibinfo {author} {\bibfnamefont {C.~M.}\
  \bibnamefont {Hirata}},\ }\href
  {{https://link.aps.org/doi/10.1103/PhysRevD.83.043513}} {\bibfield  {journal}
  {\bibinfo  {journal} {{Phys. Rev.}}\ }\textbf {\bibinfo {volume} {{D83}}},\
  \bibinfo {pages} {{043513}} (\bibinfo {year} {{2011}})},\ \Eprint
  {http://arxiv.org/abs/{arXiv:1011.3758}} {arXiv:{arXiv:1011.3758}}
  \BibitemShut {NoStop}%
\bibitem [{\citenamefont {Aghanim}\ \emph {et~al.}(2016)\citenamefont {Aghanim}
  \emph {et~al.}}]{Aghanim:2015xee}%
  \BibitemOpen
  \bibfield  {author} {\bibinfo {author} {\bibfnamefont {N.}~\bibnamefont
  {Aghanim}} \emph {et~al.} (\bibinfo {collaboration} {Planck}),\ }\href
  {\doibase 10.1051/0004-6361/201526926} {\bibfield  {journal} {\bibinfo
  {journal} {Astron. Astrophys.}\ }\textbf {\bibinfo {volume} {594}},\ \bibinfo
  {pages} {A11} (\bibinfo {year} {2016})},\ \Eprint
  {http://arxiv.org/abs/1507.02704} {arXiv:1507.02704 [astro-ph.CO]}
  \BibitemShut {NoStop}%
\bibitem [{\citenamefont {Ade}\ \emph {et~al.}(2016{\natexlab{b}})\citenamefont
  {Ade} \emph {et~al.}}]{Ade:2015zua}%
  \BibitemOpen
  \bibfield  {author} {\bibinfo {author} {\bibfnamefont {P.~A.~R.}\
  \bibnamefont {Ade}} \emph {et~al.} (\bibinfo {collaboration} {Planck}),\
  }\href {\doibase 10.1051/0004-6361/201525941} {\bibfield  {journal} {\bibinfo
   {journal} {Astron. Astrophys.}\ }\textbf {\bibinfo {volume} {594}},\
  \bibinfo {pages} {A15} (\bibinfo {year} {2016}{\natexlab{b}})},\ \Eprint
  {http://arxiv.org/abs/1502.01591} {arXiv:1502.01591 [astro-ph.CO]}
  \BibitemShut {NoStop}%
\bibitem [{\citenamefont {Beutler}\ \emph {et~al.}(2011)\citenamefont
  {Beutler}, \citenamefont {Blake}, \citenamefont {Colless}, \citenamefont
  {Jones}, \citenamefont {Staveley-Smith}, \citenamefont {Campbell},
  \citenamefont {Parker}, \citenamefont {Saunders},\ and\ \citenamefont
  {Watson}}]{Beutler:2011hx}%
  \BibitemOpen
  \bibfield  {author} {\bibinfo {author} {\bibfnamefont {F.}~\bibnamefont
  {Beutler}}, \bibinfo {author} {\bibfnamefont {C.}~\bibnamefont {Blake}},
  \bibinfo {author} {\bibfnamefont {M.}~\bibnamefont {Colless}}, \bibinfo
  {author} {\bibfnamefont {D.~H.}\ \bibnamefont {Jones}}, \bibinfo {author}
  {\bibfnamefont {L.}~\bibnamefont {Staveley-Smith}}, \bibinfo {author}
  {\bibfnamefont {L.}~\bibnamefont {Campbell}}, \bibinfo {author}
  {\bibfnamefont {Q.}~\bibnamefont {Parker}}, \bibinfo {author} {\bibfnamefont
  {W.}~\bibnamefont {Saunders}}, \ and\ \bibinfo {author} {\bibfnamefont
  {F.}~\bibnamefont {Watson}},\ }\href {\doibase
  10.1111/j.1365-2966.2011.19250.x} {\bibfield  {journal} {\bibinfo  {journal}
  {Mon. Not. Roy. Astron. Soc.}\ }\textbf {\bibinfo {volume} {416}},\ \bibinfo
  {pages} {3017} (\bibinfo {year} {2011})},\ \Eprint
  {http://arxiv.org/abs/1106.3366} {arXiv:1106.3366 [astro-ph.CO]} \BibitemShut
  {NoStop}%
\bibitem [{\citenamefont {Ross}\ \emph {et~al.}(2015)\citenamefont {Ross},
  \citenamefont {Samushia}, \citenamefont {Howlett}, \citenamefont {Percival},
  \citenamefont {Burden},\ and\ \citenamefont {Manera}}]{Ross:2014qpa}%
  \BibitemOpen
  \bibfield  {author} {\bibinfo {author} {\bibfnamefont {A.~J.}\ \bibnamefont
  {Ross}}, \bibinfo {author} {\bibfnamefont {L.}~\bibnamefont {Samushia}},
  \bibinfo {author} {\bibfnamefont {C.}~\bibnamefont {Howlett}}, \bibinfo
  {author} {\bibfnamefont {W.~J.}\ \bibnamefont {Percival}}, \bibinfo {author}
  {\bibfnamefont {A.}~\bibnamefont {Burden}}, \ and\ \bibinfo {author}
  {\bibfnamefont {M.}~\bibnamefont {Manera}},\ }\href {\doibase
  10.1093/mnras/stv154} {\bibfield  {journal} {\bibinfo  {journal} {Mon. Not.
  Roy. Astron. Soc.}\ }\textbf {\bibinfo {volume} {449}},\ \bibinfo {pages}
  {835} (\bibinfo {year} {2015})},\ \Eprint {http://arxiv.org/abs/1409.3242}
  {arXiv:1409.3242 [astro-ph.CO]} \BibitemShut {NoStop}%
\bibitem [{\citenamefont {Ata}\ \emph {et~al.}(2018)\citenamefont {Ata} \emph
  {et~al.}}]{Ata:2017dya}%
  \BibitemOpen
  \bibfield  {author} {\bibinfo {author} {\bibfnamefont {M.}~\bibnamefont
  {Ata}} \emph {et~al.},\ }\href {\doibase 10.1093/mnras/stx2630} {\bibfield
  {journal} {\bibinfo  {journal} {Mon. Not. Roy. Astron. Soc.}\ }\textbf
  {\bibinfo {volume} {473}},\ \bibinfo {pages} {4773} (\bibinfo {year}
  {2018})},\ \Eprint {http://arxiv.org/abs/1705.06373} {arXiv:1705.06373
  [astro-ph.CO]} \BibitemShut {NoStop}%
\bibitem [{\citenamefont {Bautista}\ \emph {et~al.}(2017)\citenamefont
  {Bautista} \emph {et~al.}}]{Bautista:2017zgn}%
  \BibitemOpen
  \bibfield  {author} {\bibinfo {author} {\bibfnamefont {J.~E.}\ \bibnamefont
  {Bautista}} \emph {et~al.},\ }\href {\doibase 10.1051/0004-6361/201730533}
  {\bibfield  {journal} {\bibinfo  {journal} {Astron. Astrophys.}\ }\textbf
  {\bibinfo {volume} {603}},\ \bibinfo {pages} {A12} (\bibinfo {year}
  {2017})},\ \Eprint {http://arxiv.org/abs/1702.00176} {arXiv:1702.00176
  [astro-ph.CO]} \BibitemShut {NoStop}%
\bibitem [{\citenamefont {Alam}\ \emph {et~al.}(2017)\citenamefont {Alam} \emph
  {et~al.}}]{Alam:2016hwk}%
  \BibitemOpen
  \bibfield  {author} {\bibinfo {author} {\bibfnamefont {S.}~\bibnamefont
  {Alam}} \emph {et~al.} (\bibinfo {collaboration} {BOSS}),\ }\href {\doibase
  10.1093/mnras/stx721} {\bibfield  {journal} {\bibinfo  {journal} {Mon. Not.
  Roy. Astron. Soc.}\ }\textbf {\bibinfo {volume} {470}},\ \bibinfo {pages}
  {2617} (\bibinfo {year} {2017})},\ \Eprint {http://arxiv.org/abs/1607.03155}
  {arXiv:1607.03155 [astro-ph.CO]} \BibitemShut {NoStop}%
\bibitem [{\citenamefont {Betoule}\ \emph {et~al.}(2014)\citenamefont {Betoule}
  \emph {et~al.}}]{Betoule:2014frx}%
  \BibitemOpen
  \bibfield  {author} {\bibinfo {author} {\bibfnamefont {M.}~\bibnamefont
  {Betoule}} \emph {et~al.} (\bibinfo {collaboration} {SDSS}),\ }\href
  {\doibase 10.1051/0004-6361/201423413} {\bibfield  {journal} {\bibinfo
  {journal} {Astron. Astrophys.}\ }\textbf {\bibinfo {volume} {568}},\ \bibinfo
  {pages} {A22} (\bibinfo {year} {2014})},\ \Eprint
  {http://arxiv.org/abs/1401.4064} {arXiv:1401.4064 [astro-ph.CO]} \BibitemShut
  {NoStop}%
\bibitem [{\citenamefont {Simon}\ \emph {et~al.}(2005)\citenamefont {Simon},
  \citenamefont {Verde},\ and\ \citenamefont {Jimenez}}]{Simon:2004tf}%
  \BibitemOpen
  \bibfield  {author} {\bibinfo {author} {\bibfnamefont {J.}~\bibnamefont
  {Simon}}, \bibinfo {author} {\bibfnamefont {L.}~\bibnamefont {Verde}}, \ and\
  \bibinfo {author} {\bibfnamefont {R.}~\bibnamefont {Jimenez}},\ }\href
  {\doibase 10.1103/PhysRevD.71.123001} {\bibfield  {journal} {\bibinfo
  {journal} {Phys. Rev.}\ }\textbf {\bibinfo {volume} {D71}},\ \bibinfo {pages}
  {123001} (\bibinfo {year} {2005})},\ \Eprint
  {http://arxiv.org/abs/astro-ph/0412269} {arXiv:astro-ph/0412269 [astro-ph]}
  \BibitemShut {NoStop}%
\bibitem [{\citenamefont {Stern}\ \emph {et~al.}(2010)\citenamefont {Stern},
  \citenamefont {Jimenez}, \citenamefont {Verde}, \citenamefont
  {Kamionkowski},\ and\ \citenamefont {Stanford}}]{Stern:2009ep}%
  \BibitemOpen
  \bibfield  {author} {\bibinfo {author} {\bibfnamefont {D.}~\bibnamefont
  {Stern}}, \bibinfo {author} {\bibfnamefont {R.}~\bibnamefont {Jimenez}},
  \bibinfo {author} {\bibfnamefont {L.}~\bibnamefont {Verde}}, \bibinfo
  {author} {\bibfnamefont {M.}~\bibnamefont {Kamionkowski}}, \ and\ \bibinfo
  {author} {\bibfnamefont {S.~A.}\ \bibnamefont {Stanford}},\ }\href {\doibase
  10.1088/1475-7516/2010/02/008} {\bibfield  {journal} {\bibinfo  {journal}
  {JCAP}\ }\textbf {\bibinfo {volume} {1002}},\ \bibinfo {pages} {008}
  (\bibinfo {year} {2010})},\ \Eprint {http://arxiv.org/abs/0907.3149}
  {arXiv:0907.3149 [astro-ph.CO]} \BibitemShut {NoStop}%
\bibitem [{\citenamefont {Zhang}\ \emph {et~al.}(2014)\citenamefont {Zhang},
  \citenamefont {Zhang}, \citenamefont {Yuan}, \citenamefont {Zhang},\ and\
  \citenamefont {Sun}}]{Zhang:2012mp}%
  \BibitemOpen
  \bibfield  {author} {\bibinfo {author} {\bibfnamefont {C.}~\bibnamefont
  {Zhang}}, \bibinfo {author} {\bibfnamefont {H.}~\bibnamefont {Zhang}},
  \bibinfo {author} {\bibfnamefont {S.}~\bibnamefont {Yuan}}, \bibinfo {author}
  {\bibfnamefont {T.-J.}\ \bibnamefont {Zhang}}, \ and\ \bibinfo {author}
  {\bibfnamefont {Y.-C.}\ \bibnamefont {Sun}},\ }\href {\doibase
  10.1088/1674-4527/14/10/002} {\bibfield  {journal} {\bibinfo  {journal} {Res.
  Astron. Astrophys.}\ }\textbf {\bibinfo {volume} {14}},\ \bibinfo {pages}
  {1221} (\bibinfo {year} {2014})},\ \Eprint {http://arxiv.org/abs/1207.4541}
  {arXiv:1207.4541 [astro-ph.CO]} \BibitemShut {NoStop}%
\bibitem [{\citenamefont {Moresco}\ \emph {et~al.}(2012)\citenamefont {Moresco}
  \emph {et~al.}}]{Moresco:2012jh}%
  \BibitemOpen
  \bibfield  {author} {\bibinfo {author} {\bibfnamefont {M.}~\bibnamefont
  {Moresco}} \emph {et~al.},\ }\href {\doibase 10.1088/1475-7516/2012/08/006}
  {\bibfield  {journal} {\bibinfo  {journal} {JCAP}\ }\textbf {\bibinfo
  {volume} {1208}},\ \bibinfo {pages} {006} (\bibinfo {year} {2012})},\ \Eprint
  {http://arxiv.org/abs/1201.3609} {arXiv:1201.3609 [astro-ph.CO]} \BibitemShut
  {NoStop}%
\bibitem [{\citenamefont {Moresco}(2015)}]{Moresco:2015cya}%
  \BibitemOpen
  \bibfield  {author} {\bibinfo {author} {\bibfnamefont {M.}~\bibnamefont
  {Moresco}},\ }\href {\doibase 10.1093/mnrasl/slv037} {\bibfield  {journal}
  {\bibinfo  {journal} {Mon. Not. Roy. Astron. Soc.}\ }\textbf {\bibinfo
  {volume} {450}},\ \bibinfo {pages} {L16} (\bibinfo {year} {2015})},\ \Eprint
  {http://arxiv.org/abs/1503.01116} {arXiv:1503.01116 [astro-ph.CO]}
  \BibitemShut {NoStop}%
\bibitem [{\citenamefont {Moresco}\ \emph {et~al.}(2016)\citenamefont
  {Moresco}, \citenamefont {Pozzetti}, \citenamefont {Cimatti}, \citenamefont
  {Jimenez}, \citenamefont {Maraston}, \citenamefont {Verde}, \citenamefont
  {Thomas}, \citenamefont {Citro}, \citenamefont {Tojeiro},\ and\ \citenamefont
  {Wilkinson}}]{Moresco:2016mzx}%
  \BibitemOpen
  \bibfield  {author} {\bibinfo {author} {\bibfnamefont {M.}~\bibnamefont
  {Moresco}}, \bibinfo {author} {\bibfnamefont {L.}~\bibnamefont {Pozzetti}},
  \bibinfo {author} {\bibfnamefont {A.}~\bibnamefont {Cimatti}}, \bibinfo
  {author} {\bibfnamefont {R.}~\bibnamefont {Jimenez}}, \bibinfo {author}
  {\bibfnamefont {C.}~\bibnamefont {Maraston}}, \bibinfo {author}
  {\bibfnamefont {L.}~\bibnamefont {Verde}}, \bibinfo {author} {\bibfnamefont
  {D.}~\bibnamefont {Thomas}}, \bibinfo {author} {\bibfnamefont
  {A.}~\bibnamefont {Citro}}, \bibinfo {author} {\bibfnamefont
  {R.}~\bibnamefont {Tojeiro}}, \ and\ \bibinfo {author} {\bibfnamefont
  {D.}~\bibnamefont {Wilkinson}},\ }\href {\doibase
  10.1088/1475-7516/2016/05/014} {\bibfield  {journal} {\bibinfo  {journal}
  {JCAP}\ }\textbf {\bibinfo {volume} {1605}},\ \bibinfo {pages} {014}
  (\bibinfo {year} {2016})},\ \Eprint {http://arxiv.org/abs/1601.01701}
  {arXiv:1601.01701 [astro-ph.CO]} \BibitemShut {NoStop}%
\bibitem [{\citenamefont {Cabass}\ \emph
  {et~al.}(2016{\natexlab{a}})\citenamefont {Cabass}, \citenamefont
  {Melchiorri},\ and\ \citenamefont {Pajer}}]{Cabass:2016giw}%
  \BibitemOpen
  \bibfield  {author} {\bibinfo {author} {\bibfnamefont {G.}~\bibnamefont
  {Cabass}}, \bibinfo {author} {\bibfnamefont {A.}~\bibnamefont {Melchiorri}},
  \ and\ \bibinfo {author} {\bibfnamefont {E.}~\bibnamefont {Pajer}},\ }\href
  {\doibase 10.1103/PhysRevD.93.083515} {\bibfield  {journal} {\bibinfo
  {journal} {Phys. Rev.}\ }\textbf {\bibinfo {volume} {D93}},\ \bibinfo {pages}
  {083515} (\bibinfo {year} {2016}{\natexlab{a}})},\ \Eprint
  {http://arxiv.org/abs/1602.05578} {arXiv:1602.05578 [astro-ph.CO]}
  \BibitemShut {NoStop}%
\bibitem [{\citenamefont {Cabass}\ \emph
  {et~al.}(2016{\natexlab{b}})\citenamefont {Cabass}, \citenamefont
  {di~Valentino}, \citenamefont {Melchiorri}, \citenamefont {Pajer},\ and\
  \citenamefont {Silk}}]{Cabass:2016ldu}%
  \BibitemOpen
  \bibfield  {author} {\bibinfo {author} {\bibfnamefont {G.}~\bibnamefont
  {Cabass}}, \bibinfo {author} {\bibfnamefont {E.}~\bibnamefont
  {di~Valentino}}, \bibinfo {author} {\bibfnamefont {A.}~\bibnamefont
  {Melchiorri}}, \bibinfo {author} {\bibfnamefont {E.}~\bibnamefont {Pajer}}, \
  and\ \bibinfo {author} {\bibfnamefont {J.}~\bibnamefont {Silk}},\ }\href
  {\doibase 10.1103/PhysRevD.94.023523} {\bibfield  {journal} {\bibinfo
  {journal} {Phys. Rev.}\ }\textbf {\bibinfo {volume} {D94}},\ \bibinfo {pages}
  {023523} (\bibinfo {year} {2016}{\natexlab{b}})},\ \Eprint
  {http://arxiv.org/abs/1605.00209} {arXiv:1605.00209 [astro-ph.CO]}
  \BibitemShut {NoStop}%
\bibitem [{\citenamefont {Andre}\ \emph {et~al.}(2013)\citenamefont {Andre}
  \emph {et~al.}}]{Andre:2013afa}%
  \BibitemOpen
  \bibfield  {author} {\bibinfo {author} {\bibfnamefont {P.}~\bibnamefont
  {Andre}} \emph {et~al.} (\bibinfo {collaboration} {PRISM}),\ }\href@noop {}
  {\bibfield  {journal} {\bibinfo  {journal} {ArXiv e-prints}\ } (\bibinfo
  {year} {2013})},\ \Eprint {http://arxiv.org/abs/1306.2259} {arXiv:1306.2259
  [astro-ph.CO]} \BibitemShut {NoStop}%
\bibitem [{\citenamefont {Lasky}\ \emph {et~al.}(2016)\citenamefont {Lasky}
  \emph {et~al.}}]{Lasky:2015lej}%
  \BibitemOpen
  \bibfield  {author} {\bibinfo {author} {\bibfnamefont {P.~D.}\ \bibnamefont
  {Lasky}} \emph {et~al.},\ }\href {\doibase 10.1103/PhysRevX.6.011035}
  {\bibfield  {journal} {\bibinfo  {journal} {Phys. Rev.}\ }\textbf {\bibinfo
  {volume} {X6}},\ \bibinfo {pages} {011035} (\bibinfo {year} {2016})},\
  \Eprint {http://arxiv.org/abs/1511.05994} {arXiv:1511.05994 [astro-ph.CO]}
  \BibitemShut {NoStop}%
\bibitem [{\citenamefont {Meerburg}\ \emph {et~al.}(2015)\citenamefont
  {Meerburg}, \citenamefont {Hlo\u{z}ek}, \citenamefont {Hadzhiyska},\ and\
  \citenamefont {Meyers}}]{Meerburg:2015zua}%
  \BibitemOpen
  \bibfield  {author} {\bibinfo {author} {\bibfnamefont {P.~D.}\ \bibnamefont
  {Meerburg}}, \bibinfo {author} {\bibfnamefont {R.}~\bibnamefont
  {Hlo\u{z}ek}}, \bibinfo {author} {\bibfnamefont {B.}~\bibnamefont
  {Hadzhiyska}}, \ and\ \bibinfo {author} {\bibfnamefont {J.}~\bibnamefont
  {Meyers}},\ }\href {\doibase 10.1103/PhysRevD.91.103505} {\bibfield
  {journal} {\bibinfo  {journal} {Phys. Rev.}\ }\textbf {\bibinfo {volume}
  {D91}},\ \bibinfo {pages} {103505} (\bibinfo {year} {2015})},\ \Eprint
  {http://arxiv.org/abs/1502.00302} {arXiv:1502.00302 [astro-ph.CO]}
  \BibitemShut {NoStop}%
\bibitem [{\citenamefont {Cabass}\ \emph
  {et~al.}(2016{\natexlab{c}})\citenamefont {Cabass}, \citenamefont {Pagano},
  \citenamefont {Salvati}, \citenamefont {Gerbino}, \citenamefont {Giusarma},\
  and\ \citenamefont {Melchiorri}}]{Cabass:2015jwe}%
  \BibitemOpen
  \bibfield  {author} {\bibinfo {author} {\bibfnamefont {G.}~\bibnamefont
  {Cabass}}, \bibinfo {author} {\bibfnamefont {L.}~\bibnamefont {Pagano}},
  \bibinfo {author} {\bibfnamefont {L.}~\bibnamefont {Salvati}}, \bibinfo
  {author} {\bibfnamefont {M.}~\bibnamefont {Gerbino}}, \bibinfo {author}
  {\bibfnamefont {E.}~\bibnamefont {Giusarma}}, \ and\ \bibinfo {author}
  {\bibfnamefont {A.}~\bibnamefont {Melchiorri}},\ }\href {\doibase
  10.1103/PhysRevD.93.063508} {\bibfield  {journal} {\bibinfo  {journal} {Phys.
  Rev.}\ }\textbf {\bibinfo {volume} {D93}},\ \bibinfo {pages} {063508}
  (\bibinfo {year} {2016}{\natexlab{c}})},\ \Eprint
  {http://arxiv.org/abs/1511.05146} {arXiv:1511.05146 [astro-ph.CO]}
  \BibitemShut {NoStop}%
\bibitem [{\citenamefont {Wang}\ \emph {et~al.}(2017)\citenamefont {Wang},
  \citenamefont {Cai}, \citenamefont {Liu},\ and\ \citenamefont
  {Piao}}]{Wang:2016tbj}%
  \BibitemOpen
  \bibfield  {author} {\bibinfo {author} {\bibfnamefont {Y.-T.}\ \bibnamefont
  {Wang}}, \bibinfo {author} {\bibfnamefont {Y.}~\bibnamefont {Cai}}, \bibinfo
  {author} {\bibfnamefont {Z.-G.}\ \bibnamefont {Liu}}, \ and\ \bibinfo
  {author} {\bibfnamefont {Y.-S.}\ \bibnamefont {Piao}},\ }\href {\doibase
  10.1088/1475-7516/2017/01/010} {\bibfield  {journal} {\bibinfo  {journal}
  {JCAP}\ }\textbf {\bibinfo {volume} {1701}},\ \bibinfo {pages} {010}
  (\bibinfo {year} {2017})},\ \Eprint {http://arxiv.org/abs/1612.05088}
  {arXiv:1612.05088 [astro-ph.CO]} \BibitemShut {NoStop}%
\bibitem [{\citenamefont {Guzzetti}\ \emph {et~al.}(2016)\citenamefont
  {Guzzetti}, \citenamefont {Bartolo}, \citenamefont {Liguori},\ and\
  \citenamefont {Matarrese}}]{Guzzetti:2016mkm}%
  \BibitemOpen
  \bibfield  {author} {\bibinfo {author} {\bibfnamefont {M.~C.}\ \bibnamefont
  {Guzzetti}}, \bibinfo {author} {\bibfnamefont {N.}~\bibnamefont {Bartolo}},
  \bibinfo {author} {\bibfnamefont {M.}~\bibnamefont {Liguori}}, \ and\
  \bibinfo {author} {\bibfnamefont {S.}~\bibnamefont {Matarrese}},\ }\href
  {\doibase 10.1393/ncr/i2016-10127-1} {\bibfield  {journal} {\bibinfo
  {journal} {Riv. Nuovo Cim.}\ }\textbf {\bibinfo {volume} {39}},\ \bibinfo
  {pages} {399} (\bibinfo {year} {2016})},\ \Eprint
  {http://arxiv.org/abs/1605.01615} {arXiv:1605.01615 [astro-ph.CO]}
  \BibitemShut {NoStop}%
\bibitem [{\citenamefont {Caprini}\ and\ \citenamefont
  {Figueroa}(2018)}]{Caprini:2018mtu}%
  \BibitemOpen
  \bibfield  {author} {\bibinfo {author} {\bibfnamefont {C.}~\bibnamefont
  {Caprini}}\ and\ \bibinfo {author} {\bibfnamefont {D.~G.}\ \bibnamefont
  {Figueroa}},\ }\href {\doibase 10.1088/1361-6382/aac608} {\bibfield
  {journal} {\bibinfo  {journal} {Class. Quant. Grav.}\ }\textbf {\bibinfo
  {volume} {35}},\ \bibinfo {pages} {163001} (\bibinfo {year} {2018})},\
  \Eprint {http://arxiv.org/abs/1801.04268} {arXiv:1801.04268 [astro-ph.CO]}
  \BibitemShut {NoStop}%
\end{thebibliography}%

\end{document}